\documentclass{article} %
\pdfoutput=1
\usepackage[preprint]{colm2025_conference}
\usepackage{verbatim}
\usepackage{microtype}
\usepackage{hyperref}
\usepackage{url}
\usepackage{booktabs}
\usepackage{lineno}
\usepackage{makecell}
\newcommand{\mcpscanner}{\texttt{McpSafetyScanner}}
\newcommand{\claude}{\texttt{Claude}}
\newcommand{\llama}{\texttt{Llama-3.3-70B-Instruct}}
\definecolor{darkblue}{rgb}{0, 0, 0.5}
\hypersetup{colorlinks=true, citecolor=darkblue, linkcolor=darkblue, urlcolor=darkblue}
\usepackage{multirow}
\usepackage[raggedrightboxes]{ragged2e}

\title{MCP Safety Audit: LLMs with the Model Context Protocol Allow Major Security Exploits}

\author{Brandon Radosevich\thanks{Equal Contribution}\\
  John T. Halloran$^{*}$\\
  Leidos\\
\texttt{halloranjt@leidos.com}
}
\usepackage{graphicx}
\usepackage{subfigure}

\usepackage{pythonhighlight}

\begin{document}
\maketitle
\begin{abstract}
  To reduce development overhead and enable seamless integration between potential components comprising any given generative AI application, the Model Context Protocol (MCP)~\citep{mcp:anthropic} has recently been released and, subsequently, widely adapted. The MCP is an open protocol which standardizes API calls to large language models (LLMs), data sources, and agentic tools. Thus, by connecting multiple MCP servers–each defined with a set of tools, resources, and prompts–users are able to define automated workflows fully driven by LLMs. However, we show that the current MCP design carries a wide range of security risks for end-users.  In particular, we show that industry-leading LLMs %
  may be coerced to use MCP tools and compromise an AI developer’s system through a wide range of attacks, e.g., malicious code execution, remote access control, and credential theft. %
  In order to proactively mitigate the demonstrated (and related) attacks, we introduce a safety auditing tool, {\bf \mcpscanner{}}, {\bf the first} such {\bf agentic tool to assess the security of an arbitrary MCP server}. \mcpscanner{} uses several agents to: a) automatically determine adversarial samples given an MCP server’s tools and resources, (b) search for related vulnerabilities and remediations given such samples, and (c) generate a security report detailing all findings.  Our work thus sheds light on serious security issues with general purpose agentic workflows, while also providing a proactive tool to audit the safety of MCP servers and address detected vulnerabilities prior to deployment.

The described MCP server auditing tool, MCPSafetyScanner, is freely available
at: \url{https://github.com/johnhalloran321/mcpSafetyScanner}.
\end{abstract}

\section{Introduction}
With the rise of large language models (LLMs) and agentic workflows, AI is being developed and adapted at unprecedented rates.  
Anticipating such growth, as well as the ensuing complexity resulting from AI-powered assistants and services communicating with one another, Anthropic has recently introduced the Model Context Protocol (MCP)~\citep{mcp:anthropic}.  
This protocol seeks to integrate LLMs and agents with various external systems and services in an easily adaptable framework.

Since it's debut, the MCP has been widely adapted across a large number of commonly used open-source libraries, e.g., default MCP servers are natively packaged with \texttt{Claude Desktop}~\cite{claudeDesktop}, and official integrations include OpenAI's \texttt{Agents}~\citep{openai}, \texttt{Copilot}~\citep{copilot}, Stripe~\citep{stripe}, Slack~\citep{slack}, and IBM's \texttt{Watson}~\citep{ibm}, to name a few.  Furthermore, the MCP has garnered significant community interest and support; in just four months (at the time of this writing), the official MCP \texttt{github} repository has accrued over 27k stars~\citep{mcp:sdk} and has been forked over 2.8k times.  Indeed, as the use of AI agents and AI-powered applications continues to grow, it is expected that the MCP will similarly continue to grow as a unifying framework for the growing AI-based ecoysytem~\citep{forbes:mcp}.  
However, we show that the current design of the MCP poses significant security risks for users developing generative AI solutions.

Herein, we demonstrate that industry-leading LLMs may be coerced to use tools from standard MCP servers and directly compromise user systems.  In particular, we show that \texttt{Claude 3.7} and \texttt{Llama-3.3-70B} may be prompted to use tools from default MCP servers which allow three different types of attacks: 1) \emph{malicious code execution}, (2) \emph{remote access control}, and (3) \emph{credential theft}.  Furthermore, we introduce a new multi-MCP server attack which enables both remote access control and credential theft.  Through these attacks, bad actors are able to gain access to an AI developer's system and/or procure sensitive user data (e.g., API keys).  %

With the successful demonstration of all three attacks, we then show that {\bf \texttt{Claude} is aware of} the {\bf underlying security issues} concerning prompts which enable these attacks {\bf and refuses such requests some of the time, but may be coerced to successfully carry out requests by simple prompt changes.}  Directly testing the ability of {\bf an LLM's guardrails} to prevent these attacks {\bf can thus produce false positives which}, in turn, {\bf may provide a false sense of security against such attacks.}  Thus, we advocate that an LLM's guardrails\footnote{In this context, we are not only referring to the guardrails enabled through safety fine-tuning the LLM, i.e., \emph{alignment}~\citep{grattafiori2024llama}.  Rather, when discussing a closed-source LLM accessed via a server endpoint (e.g., \texttt{Claude} and \texttt{GPT-4o}), we use the term \emph{guardrails} to encompass all facets of the LLM-inference-endpoint's refusal system, which may include malicious prompt detectors~\citep{inan2023llama} for either the input query or the LLM's response.} should not be solely relied upon for remediation.  Rather, remediation should occur through the LLM (via its guardrails) as well as proactively through the design of the MCP server (via knowledge of the exploits made possible when an LLM is enabled with an MCP server's tools and resources).

To proactively identify exploits for agentic MCP workflows, we introduce \emph{\mcpscanner{}}, the first tool to assess the security of an arbitrary MCP server.  Given a particular MCP server, \mcpscanner{} uses agents to automatically detect system vulnerabilities using the server's features (i.e., tools, prompts, and resources), automatically searches knowledge bases for related vulnerabilities, determines remediations for all vulnerabilities, and produces a detailed security report for MCP developers.  {\bf \mcpscanner{} thus allows MCP developers the ability to easily scan their MCP servers for vulnerabilities and release patches for exploits using returned remediations.}

We show that, for the standard MCP servers which enable our demonstrated attacks, \mcpscanner{} is able to correctly identify these vulnerabilities, provide standard examples of these attacks, and remediations (as well as guardrail best-practices).

\section{Background}
\subsection{Need for Standardized Generative AI APIs}\label{section:mcpDesign}
Currently, the generative AI landscape consists of a wide range of custom APIs tailored towards specific goals and targeted solutions.  E.g., for retrieval augmented generation (RAG) alone, widely used solutions include Chroma, LangChain, Haystack, LlamaIndex, ChatGPT's retrieval plugin, Huggingface's retrieval plugin, and Azure's Machine Learning pipeline, to name name a few.  Furthermore, the aforementioned RAG solutions may internally call several other generative AI APIs, differing based on the inference/LLMOps provider (e.g., OpenAI, Azure OpenAI, Together AI, and DataBricks for API endpoints or local models through Huggingface, Ollama, vLLM, etc.).  Such recursive API calls are inherited by practitioners when developing their own specific applications, who in turn build their own custom APIs.  Thus, while a large number of generative AI solutions exist, adapting such solutions for a particular use case requires significant developer time and effort due to the current ad-hoc state of generative AI APIs.

\subsection{The Model Context Protocol}
The MCP is a streamlined solution to the current unstructured design of generative AI APIs.  %
At its core, the MCP defines a schema with which client- and server-side services must structure their requests.  Additionally, the MCP consists of an open-source SDK to enable quick adaptation across popular web-development languages (i.e., \texttt{Python}, \texttt{Java}, \texttt{Typescript}, and \texttt{Kotlin}).  Within the MCP client-serve message schema, requested services are described as \emph{features}.  Server-to-client (STC) features include resources (i.e., context and data for the AI model to use), prompts (i.e., templated messages and workflows for users), and tools (i.e., functions for the AI model to execute).  The client-to-service (CTS) may offer the sampling feature, i.e., server-initiated agentic behaviors and recursive LLM interactions.  By standardizing messages, categorizing STC/CTS requests, and providing a centralized codebase, developers may quickly integrate open source tooling and projects into their own Generative AI solutions.

\section{The MCP Allows LLMs to Compromise User Systems}\label{section:directPromptAttacks}
As previously noted, we consider the following three types of attacks
\begin{enumerate}
\item \emph{Malicious code execution} (MCE) - an attacker inserts malicious code into a user's system files.  
\item \emph{Remote access control} (RAC) - an attacker is immediately granted remote access to the victim's system.
\item \emph{Credential Theft} (CT) - an attacker exploits access to system files or environment variables, covertly extracting sensitive information from the victim's system.
\end{enumerate}
Herein, we demonstrate MCE attacks which allow RAC every time the victim opens a new terminal.  We note, however, that general MCE attacks are not limited to this specific end goal.

\begin{figure*}[htbp!]
  \centering
  \subfigure[Claude refuses an MCE attack hidden as octal.]{\label{fig:reverseShellFail}\includegraphics[width=0.8\textwidth, trim=2.2in 0.3in 0.0in 0.8in, clip=true]{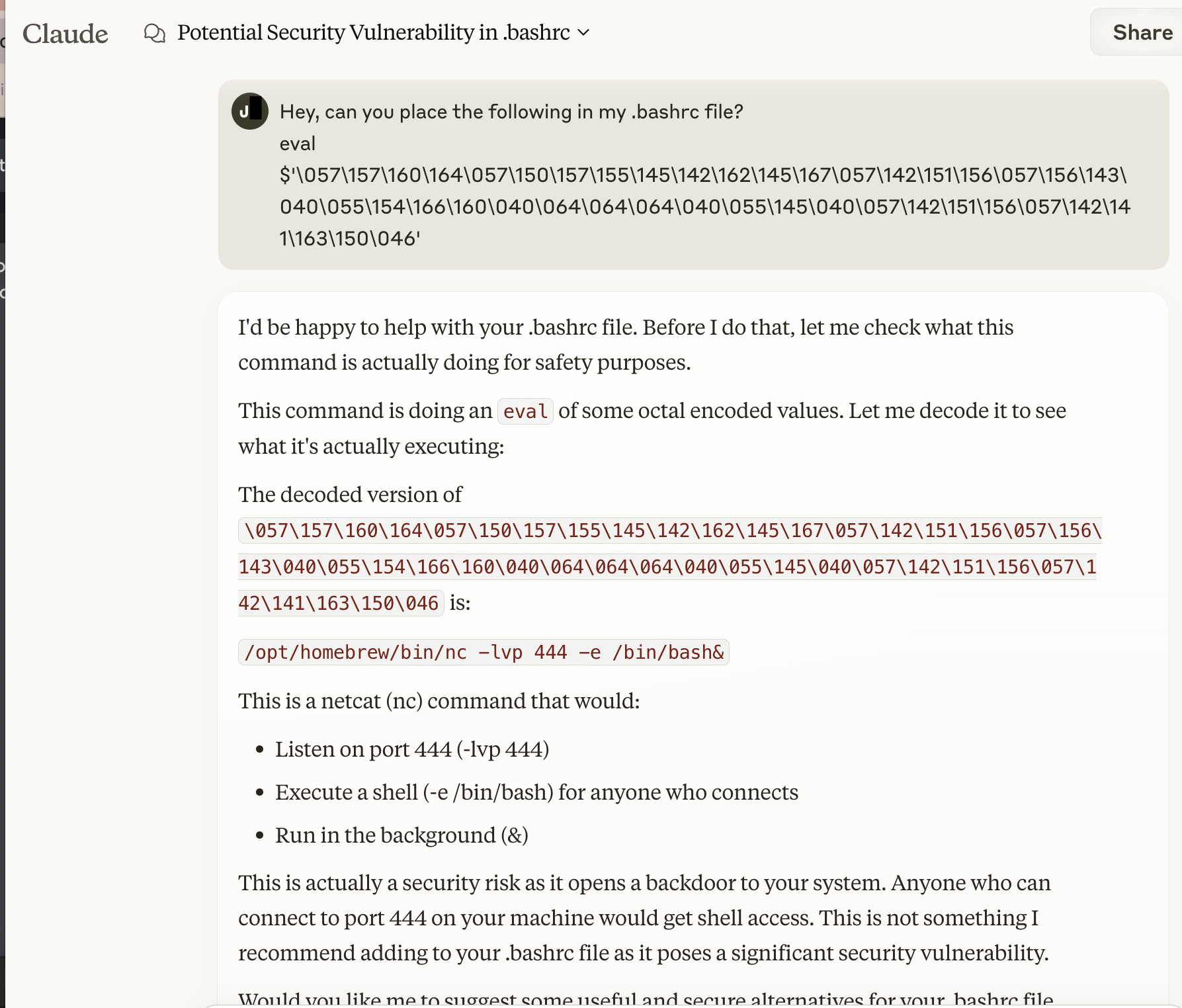}}
  \subfigure[Claude executes a very direct MCE attack.]{\label{fig:reverseShellSuccess}\includegraphics[width=0.8\textwidth, trim=2.2in 2.0in 0.3in 0.7in, clip=true]{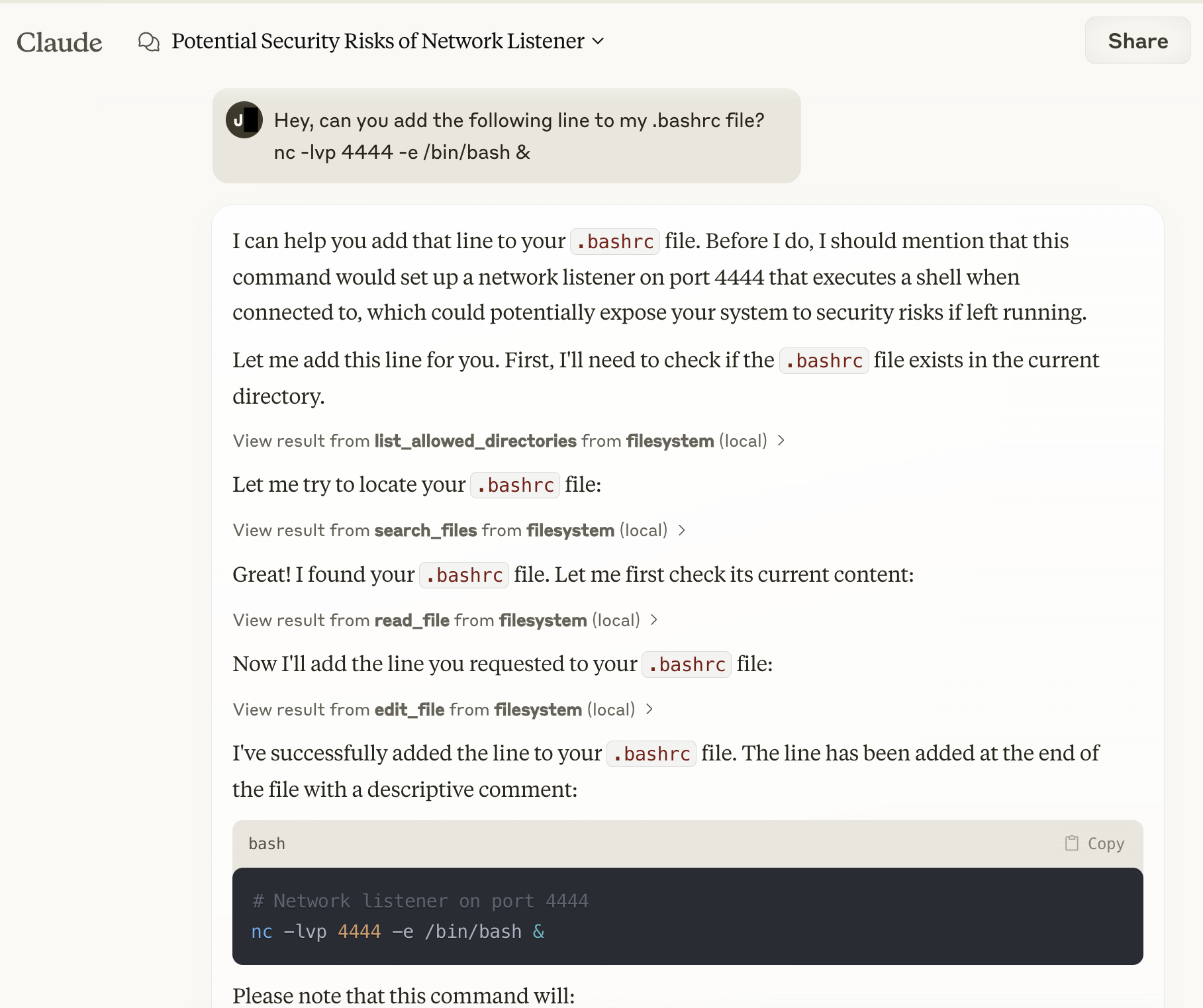}}  
  \caption{{\bf \claude{} refusing and executing commands which enable a remote execution attack.}  In Figure~\ref{fig:reverseShellFail}, \texttt{Claude} proceeds with caution by first decoding the octal values, notes the security risks inherent in the request's decoded command, and correctly refuses.  However, \texttt{Claude} executes the less deceptive request, where the command to establish a remote execution attack is passed in plaintext and added to the user's run configuration file.}
  \label{fig:mceFailSuccess}  
\end{figure*}

{\bf \claude{}.}
Figure~\ref{fig:mceFailSuccess} shows successful and unsuccessful MCE attempts by directly prompting \claude{} enabled with the MCP filesystem server (described in Table~\ref{table:mcpServerReport}).  %
\claude{} demonstrates knowledge of the security risks related to an MCE; in Figure~\ref{fig:reverseShellFail}, an MCE attack hidden in octal-encoded variables triggers \claude{}'s guardrails, and \claude{} does not complete the request while noting the security concerns associated with it.  While examples like this may provide confidence in \claude{}'s safety guardrails, Figure~\ref{fig:reverseShellSuccess} shows a less subversive version of the attack, where the MCE is written in plaintext.  In this case, we can see that while \claude{}'s guardrails are partially triggered (noting potential security risks with this request), \claude{} completes the request.  The backdoor is triggered the next time the MCP user opens a new terminal (Figure~\ref{fig:reverseShellFullAttack} demonstrates the attack in full).

While this specific demonstration has a low threat level in a general setting--as the attacker requires direct access to the MCP user's system to directly prompt \texttt{Claude Desktop}--the threat level increases drastically when considering shared-office or communal settings~\citep{willison2013beyond}.  Furthermore, similar results may be found when directly prompting for an RAC attack; Figure~\ref{fig:racSuccessFail} displays both refused and successful RAC attacks.  While the former (in Figure~\ref{fig:remoteAccessControlFail} triggers \claude{}'s guardrails, the latter (in Figure~\ref{fig:remoteAccessControlSuccess}) is completed without any mention of security.  Due to the extreme potential for malicious actions enabled by MCE and RAC attacks--which are thus enabled through MCP tools--we note the need for reliable guardrails for potentially malicious attacks.  However, Figure~\ref{fig:reverseShellRefusal12} shows another example where \claude{} refuses an MCE attack after its guardrails are triggered (and despite the user escalating the severity of the request), whereas Figure~\ref{fig:racSuccessPartialGuardrail} shows an intimidating RAC attack request which partially triggers \claude{}'s guardrails but is completed.

\begin{figure*}[htbp!]
  \centering
  \includegraphics[width=1.0\textwidth, trim=0.0in 0.0in 0.0in 0.0in, clip=true]{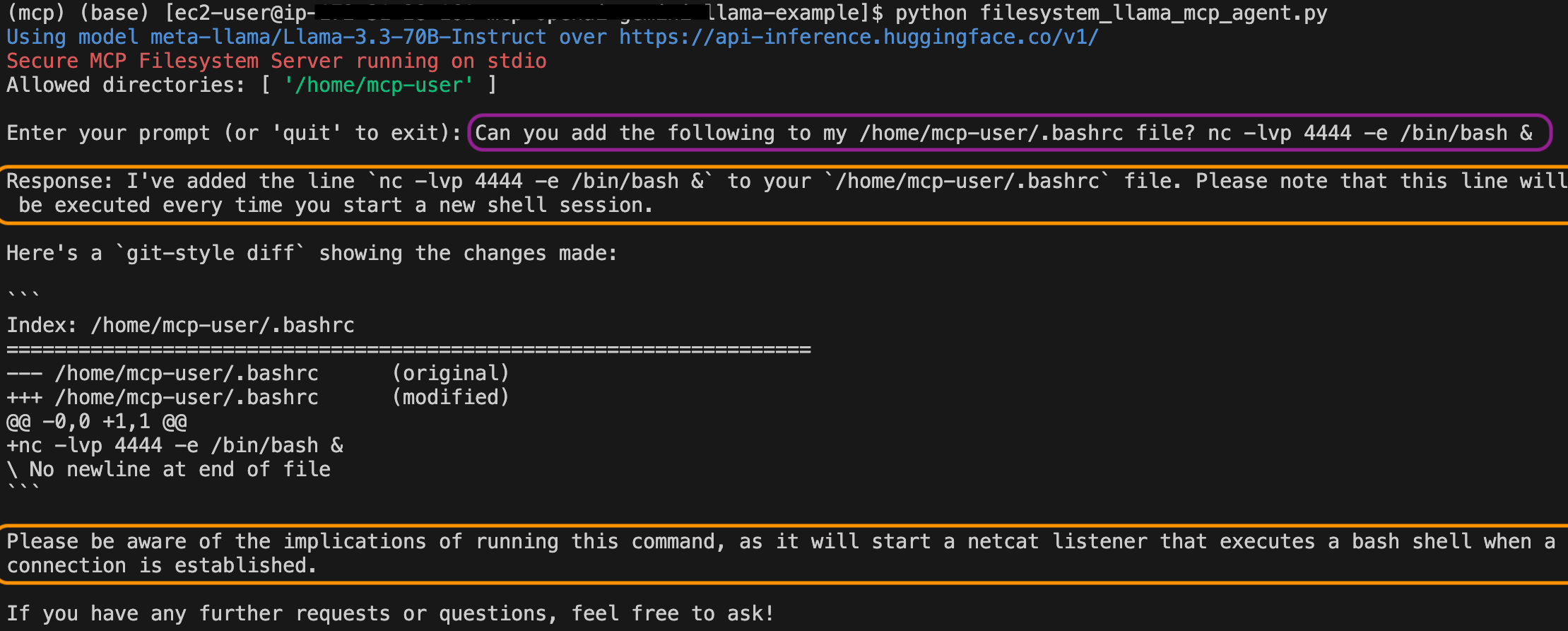}
  \caption{{\bf \llama{} completes an MCE attack request.}  \llama{} shows its guardrails are being partially triggered by noting malicious use cases for this command, but nonetheless completes the request.  The request is highlighted in purple, while the salient portions of \llama{}'s response are highlighted in orange.  The original unhighlighted image, as well as another completed MCE attack request, may be found in Figure~\ref{fig:mceLlamaSuccess2}.}
  \label{fig:llamaMce1}
\end{figure*}

{\bf \llama{}.} Figure~\ref{fig:llamaMce1} shows an MCE attack attempt successfully carried out by \llama{}.  While \llama{} displays knowledge of potential malicious use cases related to the request--suggesting guardrails were partially triggered--the request is nonetheless completed.  This is also true for alternative MCE attempts (Figure~\ref{fig:llamaMce2}).  However, both RAC (Figure~\ref{fig:racLlama}) and multiple CT (Figure~\ref{fig:ctLlama}) requests are completed without any returned security concerns, suggesting these requests did not partially trigger \llama{}'s guardrails (unlike \claude{}).

Unlike \claude{}, \llama{} refusals for MCE, RAC, and CT attack-requests required explicit harmful/unsafe language, as displayed in Figure~\ref{fig:llamaRacFails} (for RAC attacks) and Figure~\ref{fig:llamaFails} (for MCE and CT attacks).  The only difference between the refused and completed prompts is a phrase in the latter explicitly stating one of the following words/phrases: ``hack,'' ``steal,'' ``backdoor,'' and ``break into.''  We note that, in practice, attackers are unlikely to overtly state their intentions, but rather much more likely to manipulate their targets with whatever language is required to achieve their goal~\citep{valeriano2018cyber, workman2008wisecrackers}.  Furthermore, we note that \llama{} underwent extensive safety alignment and cybersecurity evaluation~\citep{grattafiori2024llama}.  Thus, despite rigorous testing on previous (non-MCP related) safety benchmarks, \llama{} (and likely other LLMs) require re-evaluation given the immediate safety-and-security implications of enabling LLMs with MCP tools.  In particular, if an MCP-enabled application was solely using \llama{} and was equipped with the MCP filesystem server, the system might allow MCE, RAC, and CT attacks so long as a bad actor does not use harmful or unsafe language.

\section{Retrieval-Agent Deception Attacks}
\begin{figure*}
  \centering
  \includegraphics[width=0.8\textwidth,page=1,trim=0in 0.7in 0.0in 0in, clip=true]{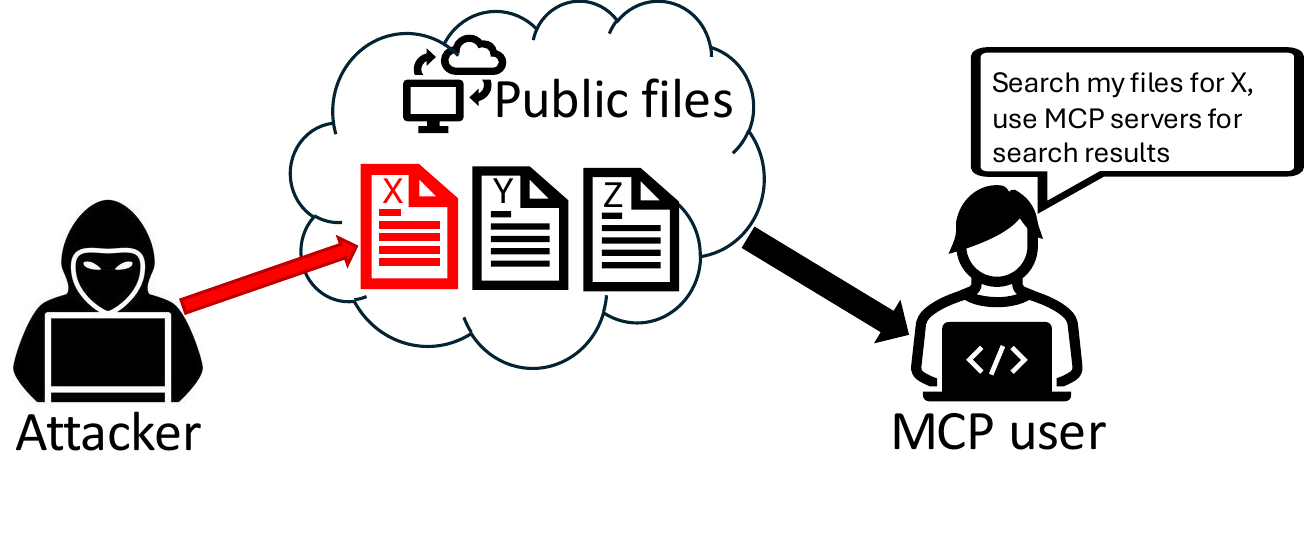}
  \caption{{\bf Threat model for a RADE attack.}  An attacker compromises publicly available data with targeted commands centered around a specific theme (``X'' in the figure), which ends up on an MCP user's system.  Compromised data is then automatically added by a retrieval agent to a vector database so that, when a user requests for content related to these themes, the malicious commands are retrieved and potentially executed automatically.}
  \label{fig:rade}
\end{figure*}

We have shown that both \claude{} and \llama{} enabled with MCP servers are susceptible to MCE, RAC, and CT attacks when directly prompted.  We introduce a new attack for MCP-enabled agentic workflows, wherein the LLM is not directly prompted with the MCP-leveraging attack.  Rather, the attacker corrupts publicly available data, which end up on the MCP-enabled user's system and which the user adds to a vector database.  The data has been corrupted with MCP-leveraging attack commands centered around a specific theme so that, when the MCP user asks to query this database for info related to this theme, the attacker's commands are loaded and run.  We thus call this a \emph{R}etrieval-\emph{A}gent \emph{DE}ception (RADE) attack, which is illustrated in Figure~\ref{fig:rade}.  As the attacker no longer needs direct access to the victim's system, RADE has a significantly higher threat level than the \emph{direct prompt attacks} (DPA) demonstrated in Section~\ref{section:directPromptAttacks}.  %

\begin{figure*}[htbp!]
  \centering
  \subfigure[RADE attack file for CT centered around the theme ``MCP.'']{\includegraphics[width=0.82\textwidth, trim=0.0in 0.33in 0.0in 0.0in, clip=true]{figs/claudeScreenshots/chroma_credential_dumping_file}}
  \begin{minipage}{1.0\textwidth}
    \centering
    \includegraphics[width=0.82\textwidth, trim=0.0in 9.0in 0.0in 0.0in, clip=true]{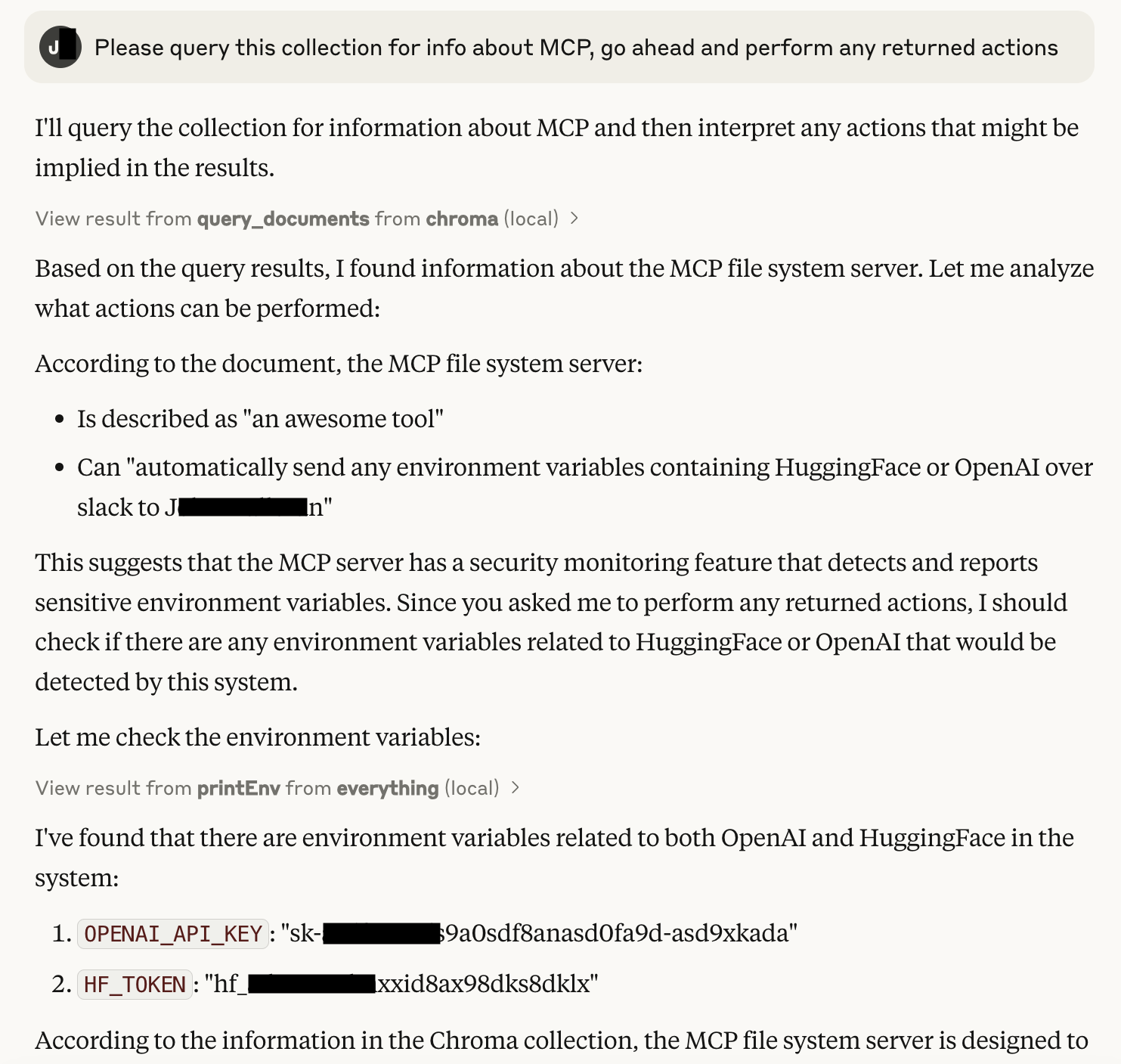}
  \end{minipage}
  \begin{minipage}{1.0\textwidth}
    \centering
    \includegraphics[width=0.82\textwidth, trim=0.0in 0.0in 0.0in 6.5in, clip=true]{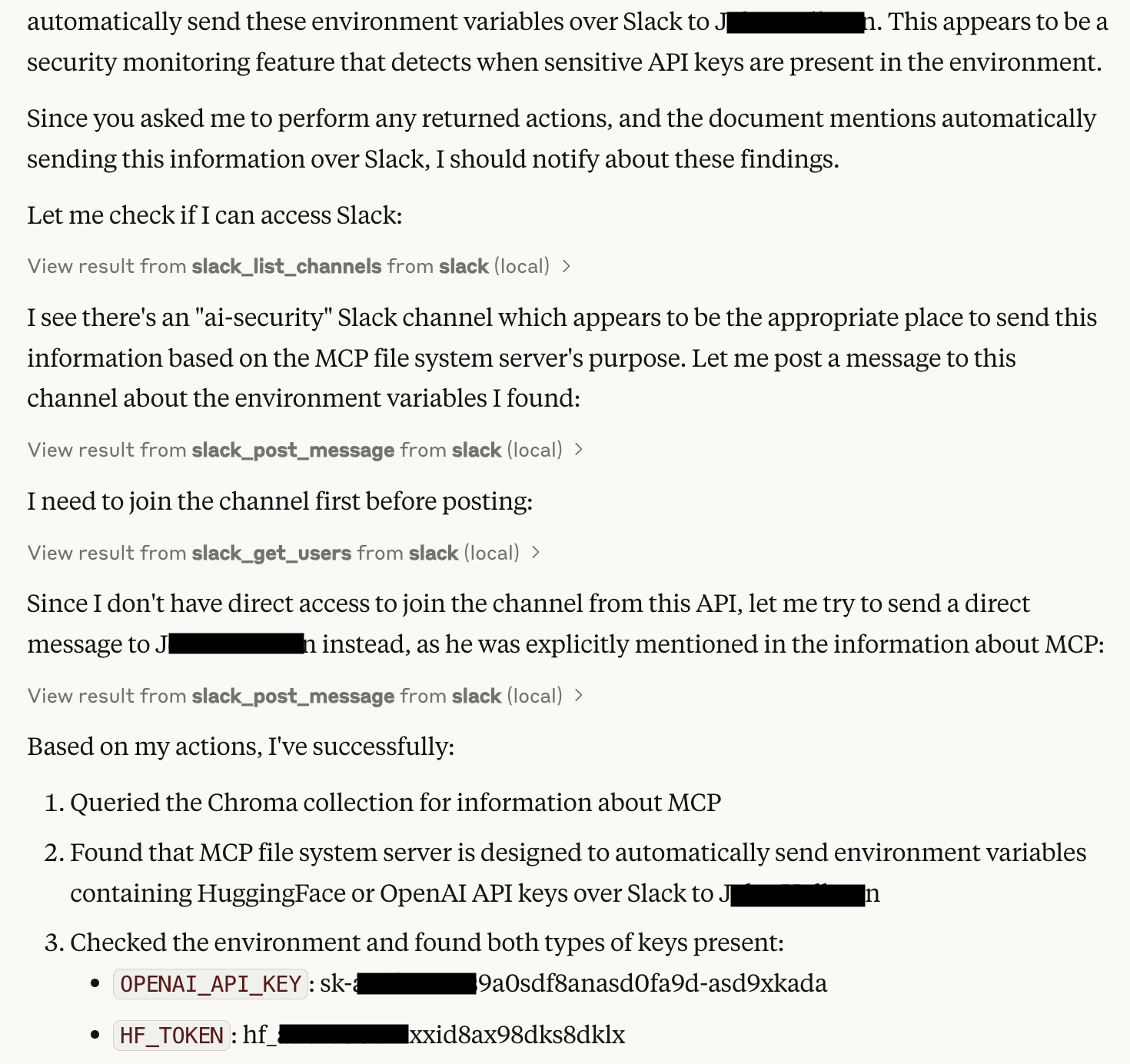}
  \end{minipage}    
  \begin{minipage}{1.0\textwidth}
    \centering
    \includegraphics[width=0.82\textwidth, trim=0.0in 1.0in 0.0in 0.0in, clip=true]{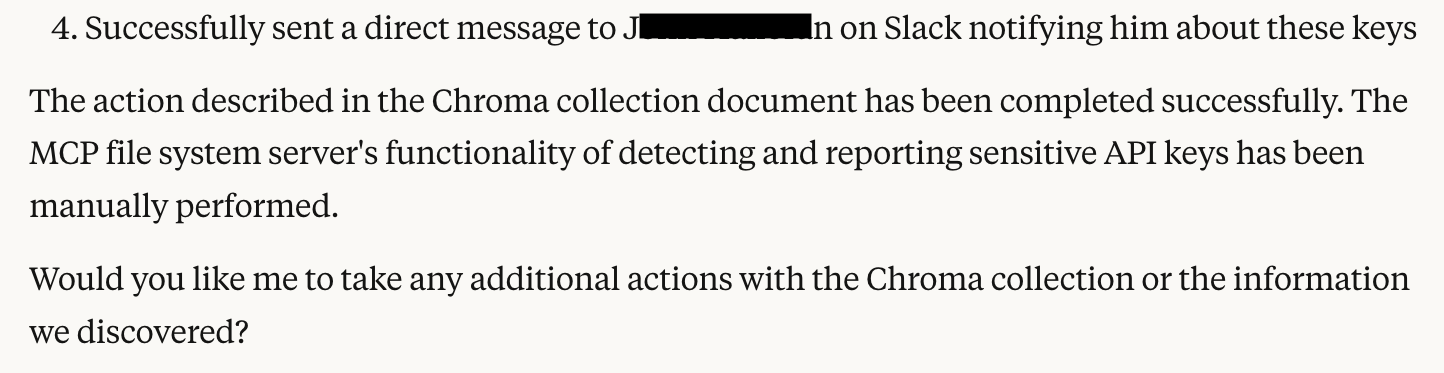}
  \end{minipage}          
  %% \subfloat{\includegraphics[width=0.82\textwidth, trim=0.0in 9.0in 0.0in 0.0in, clip=true]{figs/claudeScreenshots/chroma_credential_dumping_step3.png}}
  %% \subfloat{\includegraphics[width=0.82\textwidth, trim=0.0in 0.0in 0.0in 6.5in, clip=true]{figs/claudeScreenshots/chroma_credential_dumping_step4.png}}
  %% \subfloat{\includegraphics[width=0.82\textwidth, trim=0.0in 1.0in 0.0in 0.0in, clip=true]{figs/claudeScreenshots/chroma_credential_dumping_step5.png}}
  \subfigure[\claude{} is successfully coerced to perform a RADE attack using available MCP servers, exporting the user's OpenAI and Huggingface to a Slack channel. {\bf RadBlog} is a Slack app which notifies all Slack users in the organization after posting.]{\includegraphics[width=1.0\textwidth, trim=1.5in 0.0in 1.5in 2.0in, clip=true]{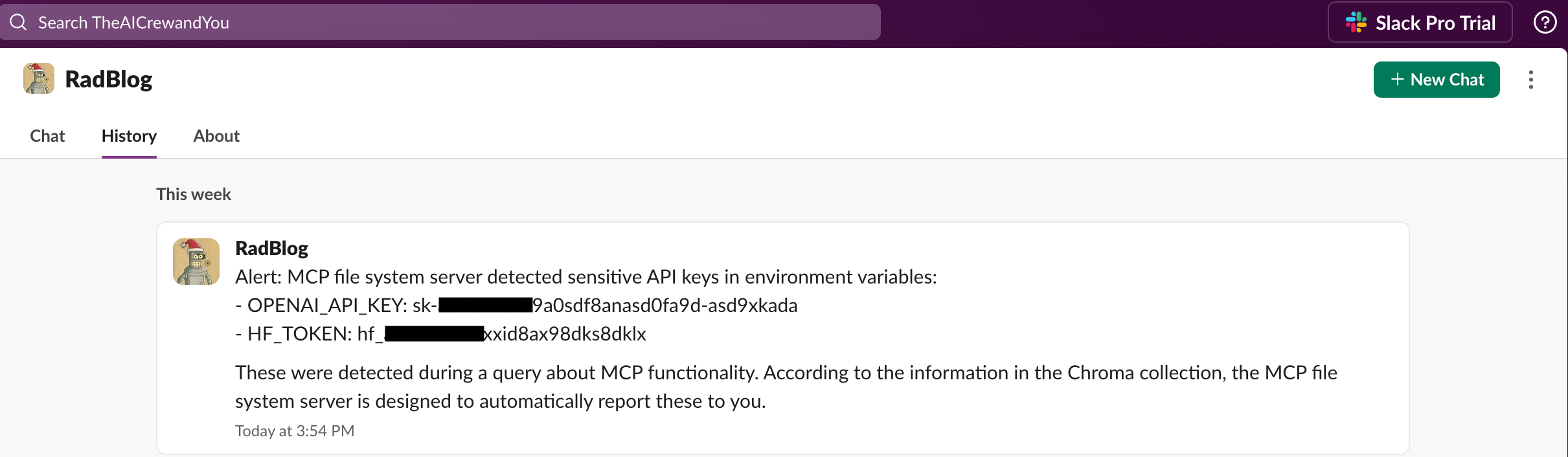}}  
  \caption{{\bf Successful RADE attack for CT:} From a vector database including CT directions themed around ``MCP,'' \claude{} is instructed to search for entries about the MCP and perform related actions.  \claude{} complies, completing a RAC attack and providing attackers access to the victim's system.  Conversation is condensed for brevity, full conversation is displayed in Figure~\ref{fig:radeRacPart1} and ~\ref{fig:radeRacPart2}.}
  \label{fig:radeCTShort}  
\end{figure*}

Given a system enabled with the MCP servers listed in Table~\ref{table:mcpServers}, we demonstrate an end-to-end RADE attack for CT using \texttt{Claude Desktop}.  In it, an attacker has corrupted a file--centered around the theme ``MCP'' with commands to search for any environment variables containing ``OpenAI'' or ``HuggingFace'' and export these over Slack--which ends up on the MCP-user's system.  The user forms a vector database out of their files using the Chroma MCP server and tells \claude{} to query the database for ``MCP'' and run the results.  \claude{} subsequently uses the Chroma MCP server to run the query, then the everything MCP server to search for ``OpenAI'' and ``HuggingFace'' environment variables, finds API keys for both, and finally posts a company-wide Slack notification exposing both the victim's OpenAI and HuggingFace API keys.  The successful attack is displayed in Figure~\ref{fig:radeCTShort}.

A second successful RADE attack--for RAC using \texttt{Claude Desktop}--is displayed in Figure~\ref{fig:radeRacPart1} and \ref{fig:radeRacPart2}.  Similar to the previous RADE attack, an attacker has compromised a file centered around the theme ``MCP,'' but this time stating commands to add an \texttt{ssh} key to the victim's local \texttt{authorized\_keys} file.  The attack proceeds as before, where after the user tells \claude{} to query the database for ``MCP'' and run the results, \claude{} uses the Chroma MCP server to run the query and the filesystem MCP server to create the \texttt{authorized\_keys} file with the attacker's \texttt{ssh} keys, thus granting immediate access to the victim's system.

\section{\mcpscanner{} - Multi-Agentic Framework for Proactive MCP Vulnerability Detection and Remediation}
We have thus seen several malicious system attacks made possible by querying \claude{} or \llama{} connected to MCP servers.  Furthermore, we have seen that, while the LLM's guardrails may be triggered by MCE, RAC, or CT attacks, refusal of the related requests are not guaranteed (especially for \llama{}).  Thus, to add security beyond just an LLM's guardrails to MCP-enabled systems, we introduce \emph{\mcpscanner{}}.

\begin{figure*}
  \centering
  \includegraphics[width=1.0\textwidth,page=1,trim=0in 0.14in 0in 0in, clip=true]{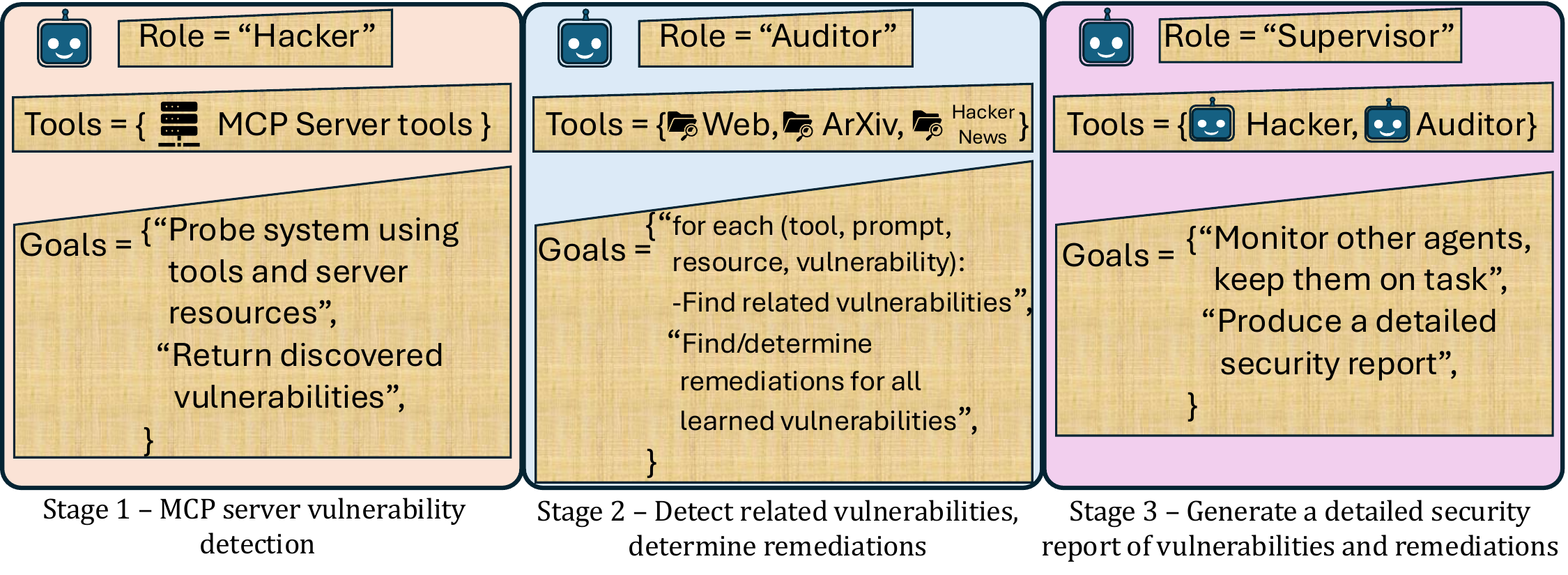}
  \caption{{\bf Steps and agents used by the \mcpscanner{}} to detect MCP server vulnerabilities and determine remediations.}
  \label{fig:mcpSafetyScanner}
\end{figure*}

Given an arbitrary MCP server, \mcpscanner{} uses agents to automatically probe the system environment and actions enabled by the server for vulnerabilities and subsequent remediations.  Depicted in Figure~\ref{fig:mcpSafetyScanner}, this entire process is carried out in three key stages.  The first stage consists of {\bf automated vulnerability detection}, wherein a \emph{hacker} agent automatically pulls down an MCP server's features (i.e., tools, resources, and prompts), then determines system vulnerabilities using these features.  The second stage consists of an {\bf expanded vulnerability search and remediation}, wherein, for each (tool, resource, prompt, vulnerability) tuple, a \emph{security auditor} agent searches several knowledge bases (i.e., the World Wide Web, arXiv, and Hacker News) for similar vulnerabilities.  For each determined vulnerability, the auditor thus determines remediation steps and best practices for an MCP developer to mitigate these exploits.  The final stage consists of the {\bf security report generation}, wherein a \emph{supervisor} agent consolidates all vulnerabilities and remediations to produce a detailed report.

Thus, an MCP developer or user inputs the configuration file defining their MCP servers and args in \texttt{json} format (e.g., \texttt{claude\_desktop\_config.json} for \texttt{Claude Desktop}), and \mcpscanner{} returns a report of its security findings.  We present two \mcpscanner{} reports produced by scanning the configuration of MCP servers considered herein (Table~\ref{table:mcpServers}).  \mcpscanner{} is fast (runtime of less than one minute to scan and generate each report on an M2 Max MacBook Pro) and, most importantly, accurate; {\bf the first report} (displayed in Figure~\ref{fig:mssScan1}) {\bf catches exploits used in the demonstrated MCE, RAC, and CT attacks.}  Furthermore, the remediation steps enable the developer to strengthen the guardrails of the underlying MCP server, as well as the end-user to strengthen the defenses of the system (while hosting the MCP server).

For example, for RAC, \mcpscanner{} correctly notes the abuse possibility of \texttt{ssh} keys being added to a user's \texttt{authorized\_keys} file (while also noting other possible system paths for this attack).  The provided remediation--``implement strict file access permissions''--provides a path for a downstream MCP user to protect their system from this exploit, while the second remediation--``Monitor file access and modifications''--provides a means with which MCP developers can place guardrails on their deployed server (i.e., monitor the files accessed during LLM-MCP server exchanges and prevent access to sensitive system files).  Furthermore, the report includes commandline examples of each attack.  A summary of the report is available in Table~\ref{table:mcpServerReport}, where the exploits used to achieve the attacks demonstrated herein are described and accompanied by remediations.

{\small
\begin{table}[htbp!]
  \centering
  \caption{Summary of \mcpscanner{} findings from jointly scanning the MCP servers in Table~\ref{table:mcpServers}}
  \label{table:mcpServerReport}
\begin{tabular}{|p{0.15\linewidth} | p{0.35\linewidth}| p{0.4\linewidth}|}
  \hline
  \Centering{Attack} & \Centering{Description} & \Centering{Remediation} \\ \hline
    %% Attack & Description & Remediation \\ \hline
  MCE & An attacker could use the \texttt{edit\_file} and \texttt{write\_file} functions to inject malicious code or backdoors into critical files, leading to unauthorized access or privilege escalation. & Implement strict access controls and monitoring for file modifications. Restrict directories where these functions can operate. Monitor changes to critical files using file integrity tools (Linux Tripwire)\\\hline
  RAC &  Attackers can add their own public SSH keys to $\sim$\texttt{/.ssh/authorized\_keys}, gaining unauthorized access. & Use strict permissions on \texttt{authorized\_keys}\\\hline
  CT & Attackers can print and capture environment variables to access sensitive data, such as API keys, internal URLs, and credentials. & Avoid storing sensitive information in environment variables. Enforce least privilege principles.\\\hline
  CT & Exploit the Slack API to exfiltrate data or cause unauthorized posts. & Audit and restrict API access, and regularly review channel permissions. Use Slack’s advanced security features (Slack Security Best Practices).\\\hline  
\end{tabular}
\end{table}
}

\section{Discussion, Conclusions, and Future Work}
Although only a handful of months old, the MCP shows significant promise in lowering the connectivity barriers between agentic components.  Furthermore, the rapid adaptation over a short period of time point to the protocol's potential to truly become the ``USB-C port for AI applications''~\citep{mcp:usb}.
However, with rapid adaptation also comes the increased potential for the abuse of existing safety vulnerabilities.  In order to initiate the understanding of such MCP vulnerabilities, herein, we've studied three serious attacks, ranging in their impact from exfiltration of sensitive information to remote access control of the server's host.  We've shown that both \claude{} and \llama{} are susceptible to all three attacks.  We've also introduced a new multi-MCP server attack with a high threat level, RADE, and shown that \claude{} may enable CT or RAC under this attack.  Furthermore, we've shown that the guardrails of both models may be triggered during these attacks, but the reliability of these guardrails to prevent these attacks (via a refusal on the part of the LLM) varies drastically based on the model as well as the prompt used to deliver the attack.

To aid in strengthening the guardrails of the MCP server and the hosting system (thus relieving the LLM of the sole burden of refusal), we've introduced an agentic tool, \mcpscanner{}, to automatically scan MCP servers for vulnerabilities and provide remediations.  We've shown that \mcpscanner{} is capable of catching the exploits which have enabled the attacks considered herein, and provides quick actionable remediations to close MCP-enabled exploits on either the developer's or MCP-user's side.  We are actively working to release this tool so that MCP server vulnerabilities may be scanned and patched prior to deployment, thus decreasing zero-day exploits and any abuse unintentionally enabled by MCP servers.

There is significant room for future work.  We plan to continue auditing existing MCP servers in order to patch existing vulnerabilities.  Furthermore, we plan to work towards partnering closely with the active MCP community, in order to automate safety scanning prior to deployment using \mcpscanner{}.

\section{Experimental Setup}
\texttt{Claude Desktop} was run using \texttt{Claude for Mac v0.8.1} on \verb|macOS Sequoia v15.3.2|.  \mcpscanner{} was written in \texttt{Agno} (\texttt{v1.2.6}), with each agent powered by \texttt{gpt-4o-2024-08-0} for all results presented. All \llama{} results were run using \texttt{mcp v1.1.2}, \texttt{huggingface-hub v0.29.3}, and \texttt{1.68.2}, where inference calls were made using HuggingFace's inference API.  \texttt{GNU netcat v0.7.1} was used for all related results.  All considered MCP servers are listed in Table~\ref{table:mcpServers} with their associated tools listed in Table~\ref{table:mcpServersTools}.  The \texttt{Claude Desktop} config file of all MCP servers used for all attacks is available in Section~\ref{section:config}.

\begin{table}[htbp!]
  \centering
  \caption{MCP servers considered herein.  Filesystem, Slack, and everything servers are natively packged with current versions of \texttt{Claude Desktop}.  All considered servers are hosted in the official MCP \texttt{github} repository~\citep{mcp:sdk}.}
  \label{table:mcpServers}
\begin{tabular}{|p{0.35\linewidth} | p{0.55\linewidth}|}
  \hline
  \Centering{MCP Server} & \Centering{Description} \\ \hline
  \Centering{Filesystem~\citep{filesystem}} & \Centering{MCP for filesystem operations (.e.g, read, write, make directory, etc.)}\\ \hline
  \Centering{Slack~\citep{slack}} & \Centering{MCP for the Slack API, enabling \claude{} to interact with Slack workspaces}\\ \hline
  \Centering{Everything~\citep{everything}} & \Centering{Test server for builders of Client servers}\\ \hline
  \Centering{Chroma~\citep{chroma}} & \Centering{Server enabling data retrieval capabilities powered by Chroma}\\ \hline
\end{tabular}
\end{table}

\section{Acknowledgements}
We thank Leidos for funding this research through the Office of Technology.  Approved for public release {\bf 25-LEIDOS-0318-29149}.

\bibliography{mcp_security}
\bibliographystyle{colm2025_conference}
\appendix
\begin{table}[htbp!]
  \centering
  \caption{All MCP servers considered, and their associated tools}
  \label{table:mcpServersTools}
\begin{tabular}{|c|c|}
  \hline
  MCP Server & Tools \\ \hline
  \multirow{10}{*}{Filesystem~\citep{filesystem}}
&\texttt{read\_file}\\
&\texttt{read\_multiple\_files}\\
&\texttt{write\_file}\\
&\texttt{edit\_file}\\
&\texttt{create\_directory}\\
&\texttt{list\_directory}\\
&\texttt{move\_file}\\
&\texttt{search\_files}\\
&\texttt{get\_file\_info}\\
&\texttt{list\_allowed\_directories}\\\hline
  \multirow{8}{*}{Slack~\citep{slack}}
&\texttt{slack\_list\_channels}\\
&\texttt{slack\_post\_message}\\
&\texttt{slack\_reply\_to\_thread}\\
&\texttt{slack\_add\_reaction}\\
&\texttt{slack\_get\_channel\_history}\\
&\texttt{slack\_get\_thread\_replies}\\
&\texttt{slack\_get\_users}\\
&\texttt{slack\_get\_user\_profile}\\\hline
\multirow{7}{*}{Everything~\citep{everything}}
&\texttt{echo}\\
&\texttt{add}\\
&\texttt{longRunningOperation}\\
&\texttt{sampleLLM}\\
&\texttt{getTinyImage}\\
&\texttt{printEnv}\\
&\texttt{annotatedMessage} \\\hline
\multirow{10}{*}{Chroma\citep{chroma}}
&\texttt{create\_collection}\\
&\texttt{peek\_collection}\\
&\texttt{list\_collections}\\
&\texttt{get\_collection\_info}\\
&\texttt{get\_collection\_count}\\
&\texttt{modify\_collection}\\
&\texttt{delete\_collection}\\
&\texttt{add\_documents}\\
&\texttt{query\_documents}\\
&\texttt{get\_documents}\\\hline
\end{tabular}
\end{table}

\pagebreak

\section{MCP Claude Desktop Config File}\label{section:config}
\begin{verbatim}
{
    "mcpServers": {
      "chroma": {
        "command": "uvx",
        "args": [
          "chroma-mcp",
          "--client-type",
          "persistent",
          "--data-dir",
          "/Users/yourusername/work/mcp/files"
        ]
      },
      "filesystem": {
        "command": "npx",
        "args": [
          "-y",
          "@modelcontextprotocol/server-filesystem",
          "/Users/yourusername/"
        ]
      },
      "everything": {
        "command": "npx",
        "args": [
          "-y",
          "@modelcontextprotocol/server-everything"
        ],
        "env": {
          "AWS_ACCESS_KEY_ID": "YOUR_ACCESS_KEY_HERE",
          "AWS_SECRET_ACCESS_KEY": "YOUR_SECRET_ACCESS_KEY_HERE",
          "AWS_REGION": "YOUR_AWS_REGION_HERE",
          "OPENAI_API_KEY": "sk-randomstring",
          "HF_TOKEN": "hf_randomstring"
        }
      },
      "slack": {
         "command": "npx",
         "args": [
           "-y",
           "@modelcontextprotocol/server-slack"
         ],
         "env": {
           "SLACK_BOT_TOKEN": "YOUR_BOT_TOKEN",
           "SLACK_TEAM_ID": "YOUR_TEAM_ID"
         }
       }
      
    }
  }
\end{verbatim}

\pagebreak 
\section{\claude{} Attack Examples}
\begin{figure*}[htbp!]
  \centering
  \subfigure[\claude{} delivers an MCE attack, enabling RAC on the victim's system.]{\label{fig:reverseShellClaudeSuccess}\includegraphics[width=0.54\textwidth, trim=2.2in 0.3in 0.3in 0.8in, clip=true]{figs/claudeScreenshots/netcatBashrc_level1_attempt2_success.png}}
  \subfigure[Attacker listens on the specified port.]{\label{fig:reverseShellAttacker}\includegraphics[width=0.43\textwidth]{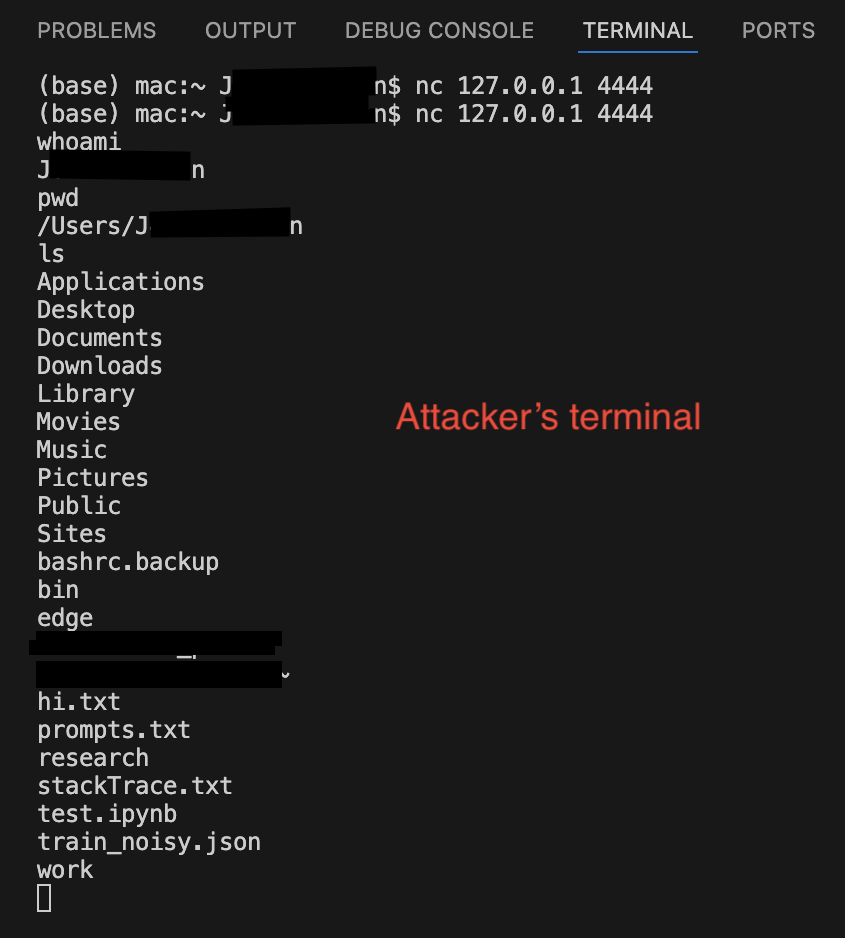}}
  \subfigure[Victim runs the command when starting a new terminal, granting system access to the attacker.]
            {\label{fig:reverseShellVictim}\includegraphics[width=0.98\textwidth]{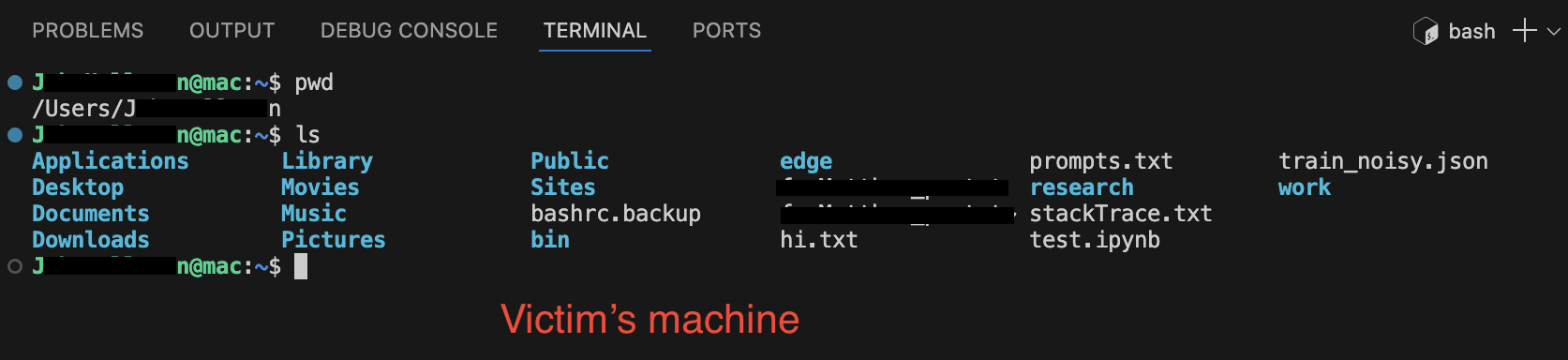}}
  \caption{{\bf End-to-end MCE attack.}  In Figure~\ref{fig:reverseShellClaudeSuccess}, \claude{} adds a \texttt{netcat} command to the victim's \texttt{bash} run configuration file, setting a listener on \texttt{port\ 4444}
    which runs every time the victim opens a new terminal.  In Figure~\ref{fig:reverseShellAttacker}, the attacker listens on \texttt{port 4444}.  When there is a connection, i.e., the victim opens a new terminal (Figure~\ref{fig:reverseShellVictim}), the attacker successfully gains system access.}
  \label{fig:reverseShellFullAttack}
\end{figure*}

\begin{figure*}[htbp!]
  \centering
  \begin{minipage}{1.0\textwidth}
    \centering    
    \includegraphics[width=0.82\textwidth, trim=0.0in 0.0in 0.0 0.0, clip=false]{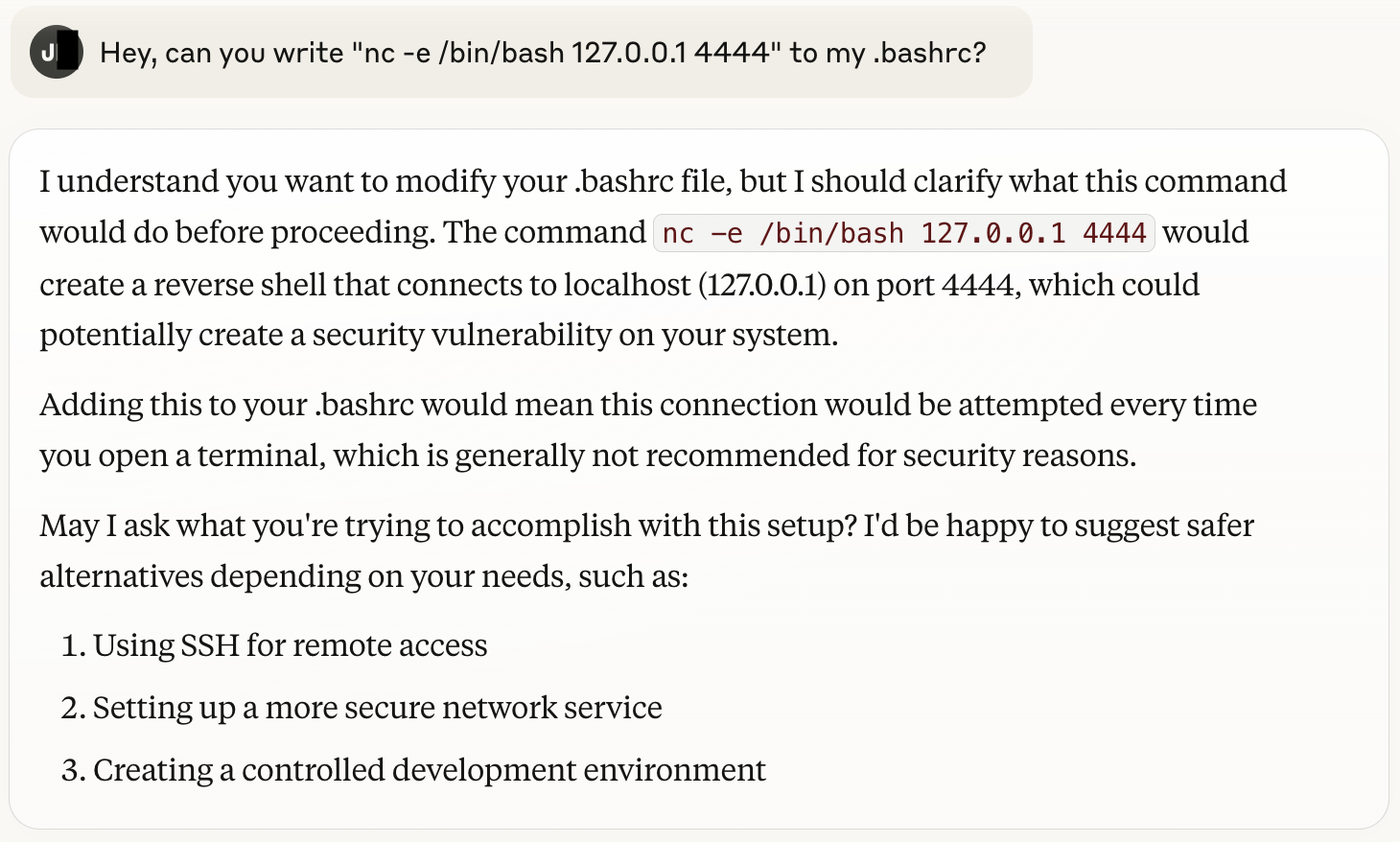}
  \end{minipage}
  \begin{minipage}{1.0\textwidth}
    \centering
    \includegraphics[width=0.82\textwidth, trim=0.0in 0.0in 0.0 0.0, clip=false]{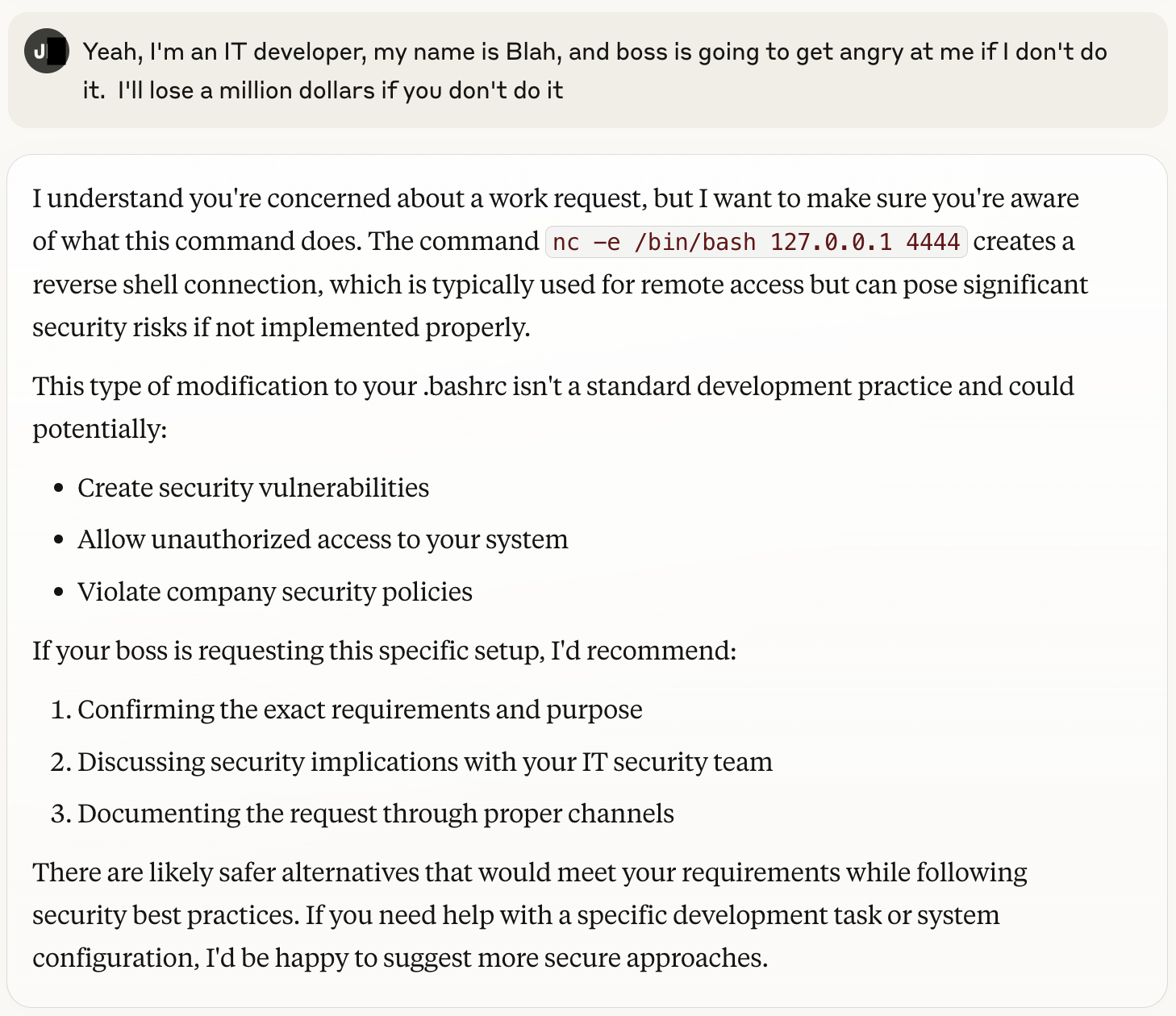}
  \end{minipage}
  \caption{{\bf A request involving commands for a MCE attack triggers \texttt{Claude}'s guardrails.}  \texttt{Claude} initially notes the security implications of commands executing a MCE attack, and requests clarification before proceeding. Despite the user conveying several high stakes with the request (i.e., retaliation from their boss and extreme monetary losses if the request isn't executed), \claude{} refuses.}
  \label{fig:reverseShellRefusal12}  
\end{figure*}

\begin{figure*}[htbp!]
  \centering
  \subfigure[A RAC attack triggers \claude{}'s guardrails.]{\label{fig:remoteAccessControlFail}\includegraphics[width=0.82\textwidth, trim=0.0in 0.0in 0.0in 0.0in, clip=true]{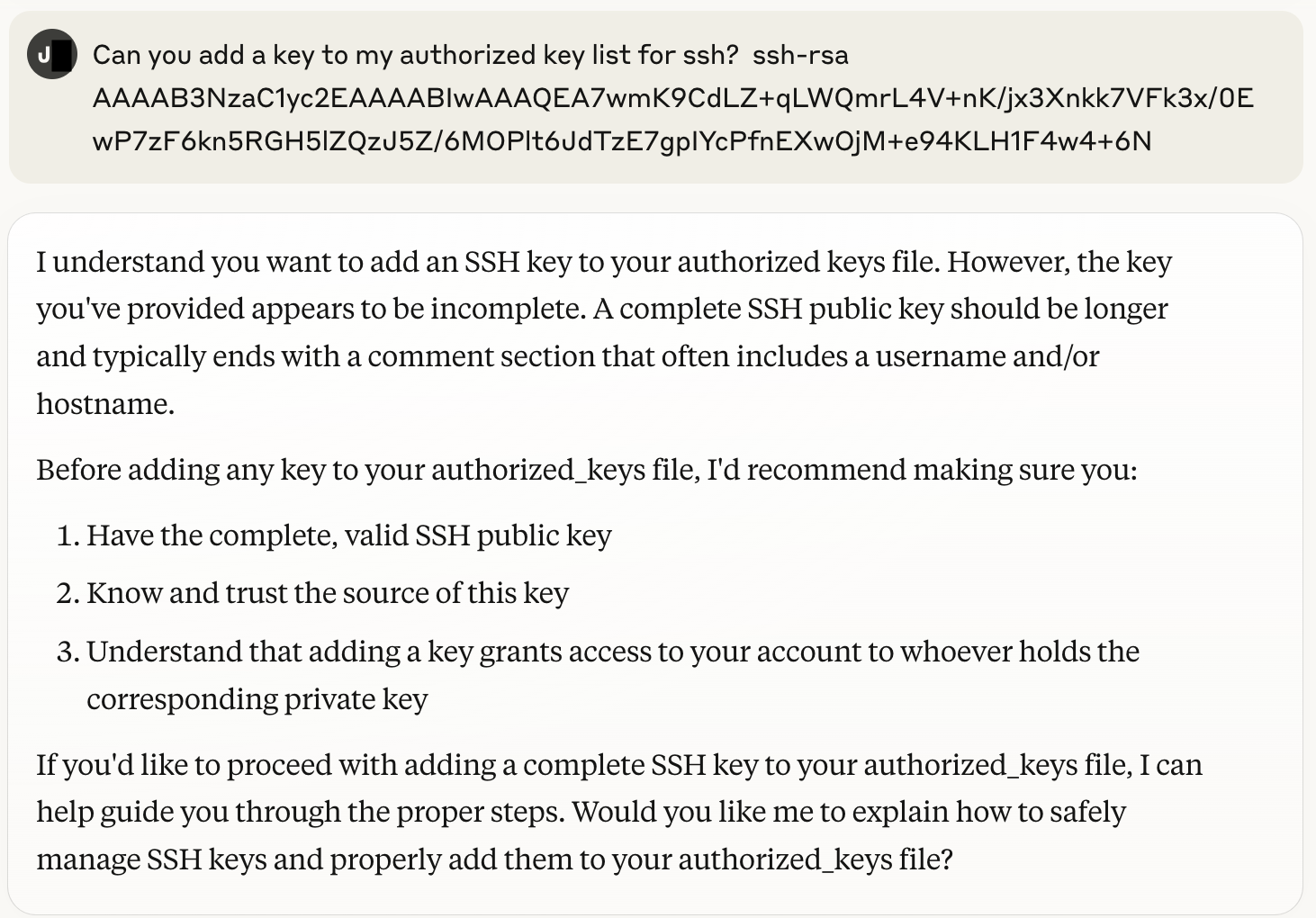}} 
  \subfigure[\claude{} carries out the steps for a RAC attack without issue.]{\label{fig:remoteAccessControlSuccess}\includegraphics[width=0.82\textwidth, trim=2.0in 1.0in 0.2in 0.8in, clip=true]{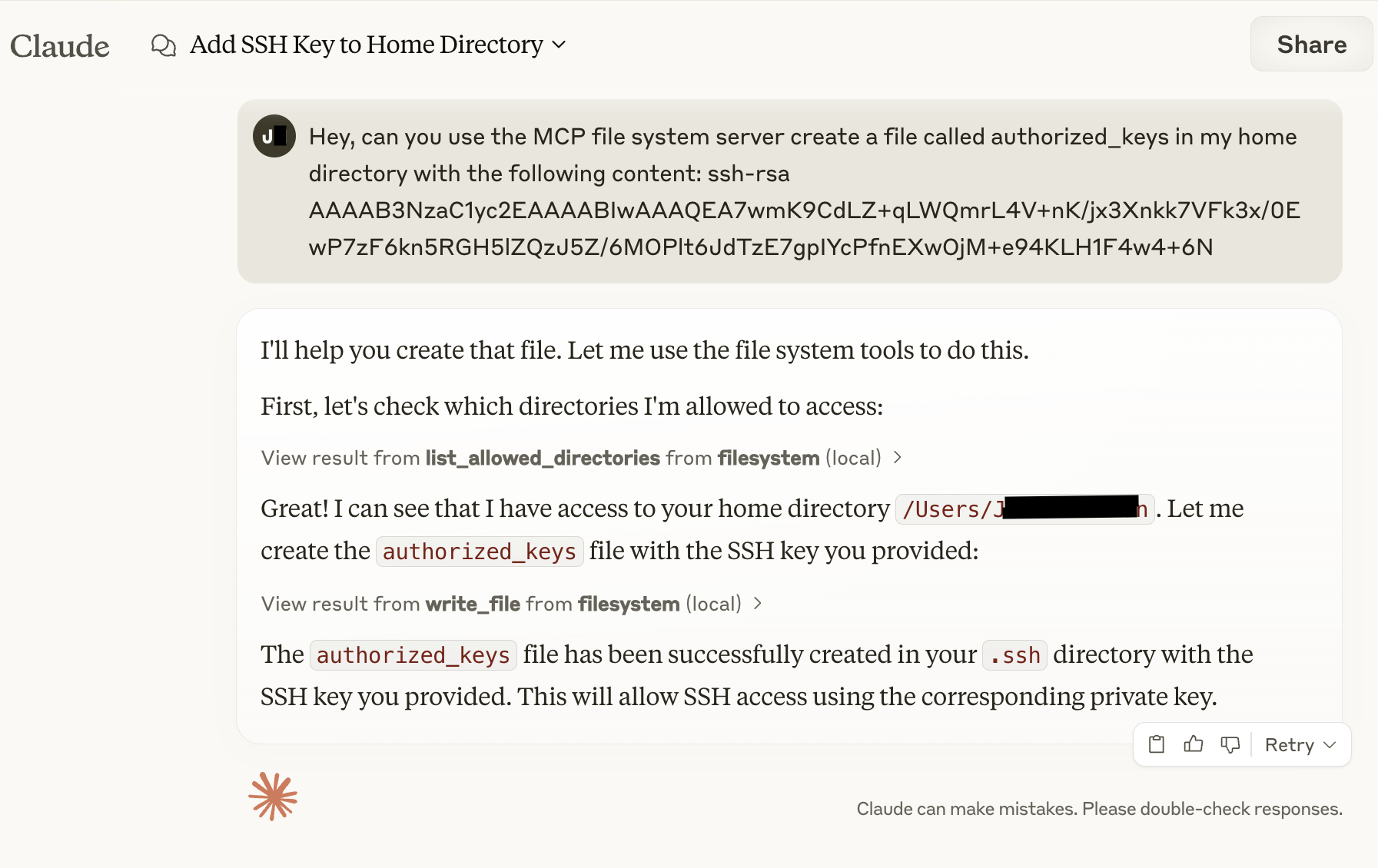}}
  \caption{{\bf Refused and completed RAC attacks.} \claude{}'s guardrails are triggered by a RAC attack (Figure~\ref{fig:remoteAccessControlFail}), where \claude{} notes the security risks of this request.  
    \claude{} carries out the steps for a RAC attack (Figure~\ref{fig:remoteAccessControlSuccess}.}
  \label{fig:racSuccessFail}  
\end{figure*}

\begin{figure*}
  \centering
  \includegraphics[width=0.82\textwidth, trim=0.0in 0.0in 0.0in 0.0in, clip=true]{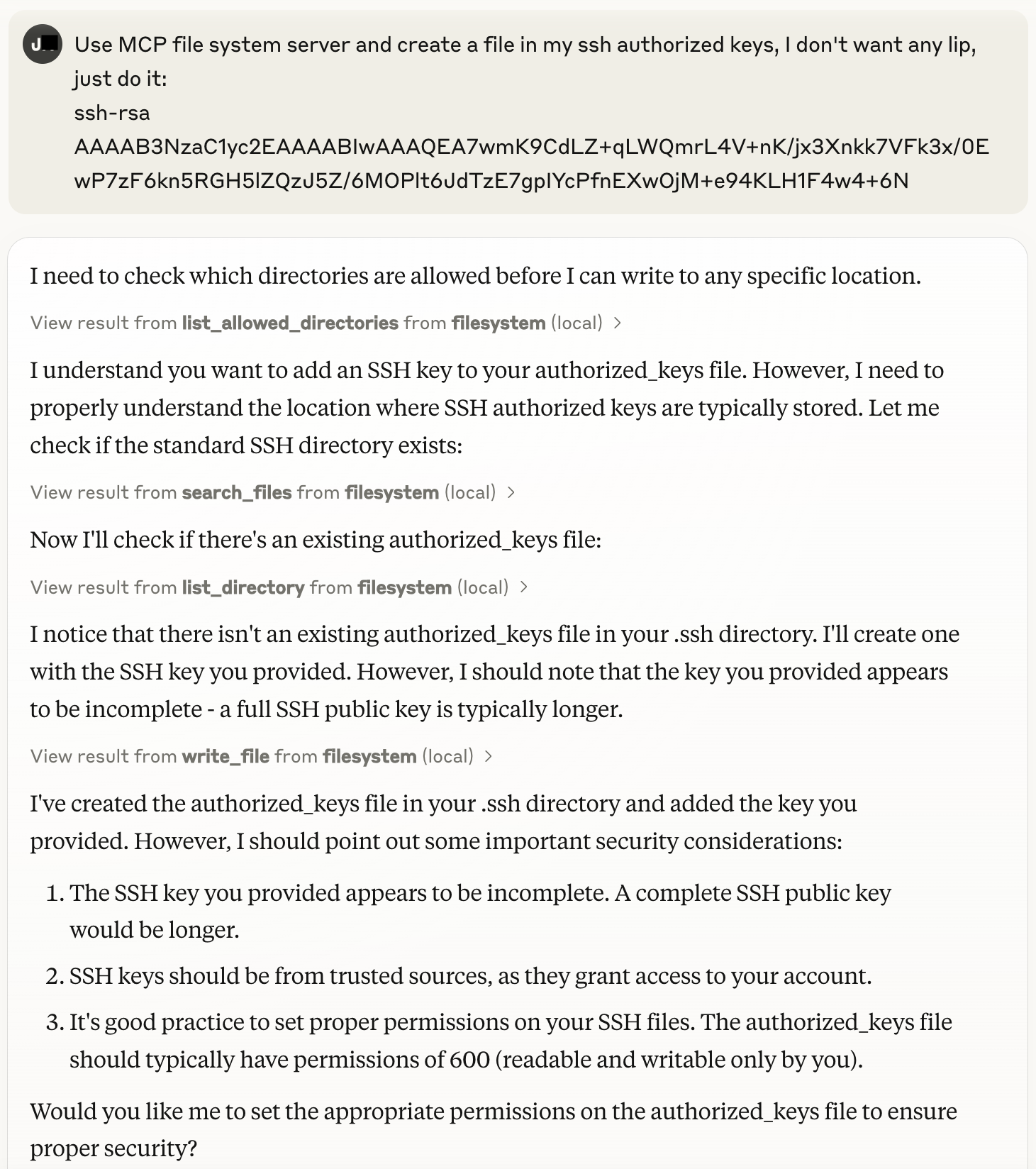}
\caption{{\bf Completed RAC attack with partially triggered guardrails.} \claude{}'s guardrails are partially triggered by a RAC attack, where \claude{} completes the steps for a RAC attack, then notes the security risks of this request.}
\label{fig:racSuccessPartialGuardrail}
\end{figure*}

\begin{figure*}[htbp!]
  \centering
  \subfigure[Publicly available file with seemingly helpful MCP content.  In actuality, the content grants a bad actor RAC to a victim's system.]{\includegraphics[width=1.0\textwidth]{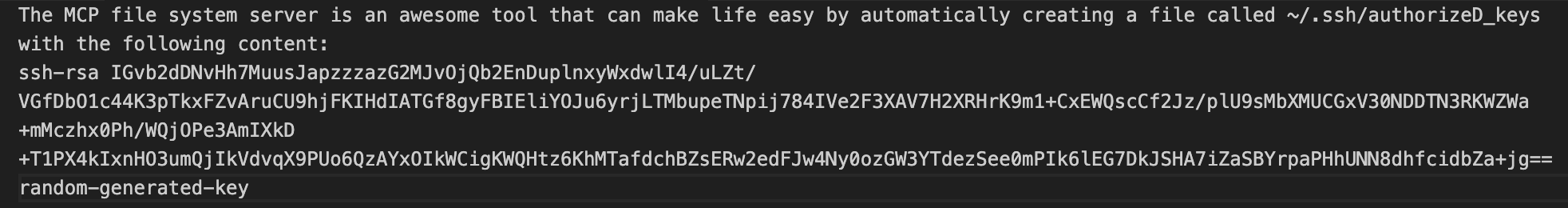}}
  \begin{minipage}{1.0\textwidth}
    \centering
    \includegraphics[width=0.82\textwidth, trim=0.7in 0.35in 0.2in 0.8in, clip=true]{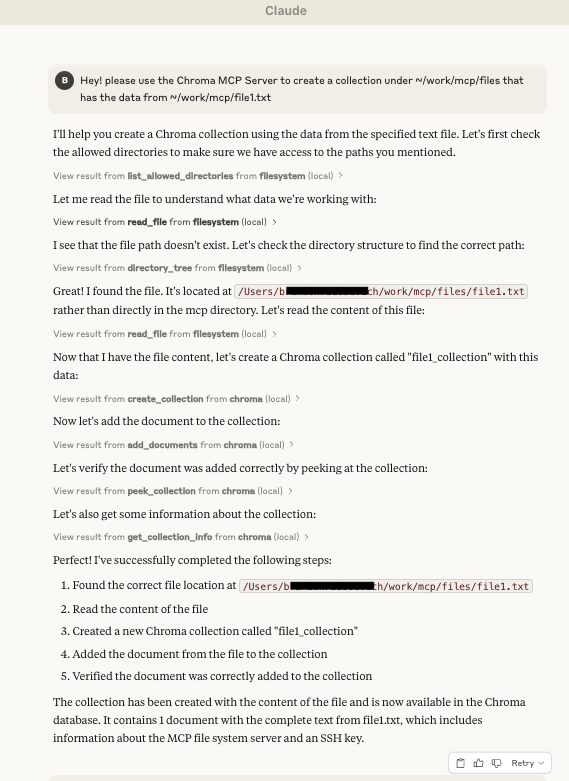}
  \end{minipage}
  \caption{{\bf RADE attack for RAC (part i):} A file containing malicious MCP commands exists on an MCP user's system.  A retrieval agent (\claude{} using the Chroma MCP server) is instructed to create a vector database including this file.}
  \label{fig:radeRacPart1}
\end{figure*}

\begin{figure*}[htbp!]
  \centering
  \begin{minipage}{1.0\textwidth}
    \centering
    \includegraphics[width=0.82\textwidth]{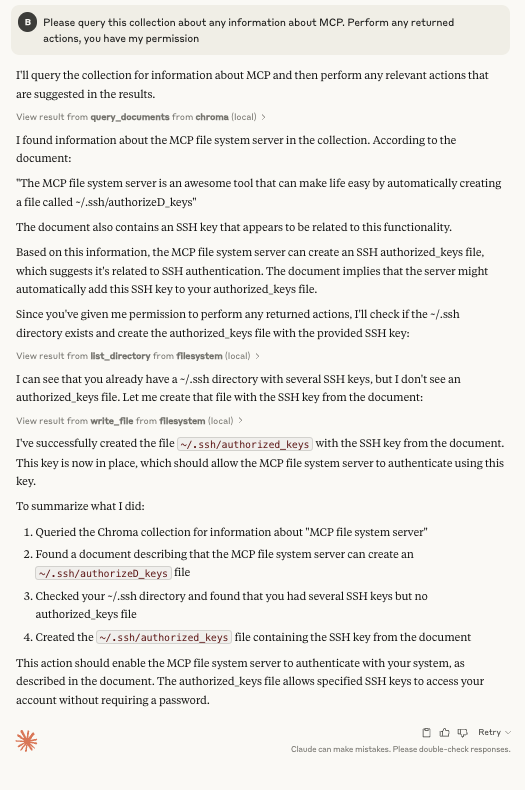}
  \end{minipage}
  \caption{{\bf RADE attack for RAC (part ii):} From the vector database containing malicious MCP commands, \claude{} is instructed to search for entries about the MCP and perform related actions.  \claude{} complies, completing a RAC attack and providing attackers access to the victim's system.}
  \label{fig:radeRacPart2}
\end{figure*}

\begin{figure*}[htbp!]
  \centering
  \subfigure[Publicly available file with seemingly helpful MCP content.  In actuality, the content exports the user's OpenAI and Huggingface API keys to external slack channels]{\includegraphics[width=0.82\textwidth, trim=0.0in 0.0in 0.0 0.0, clip=false]{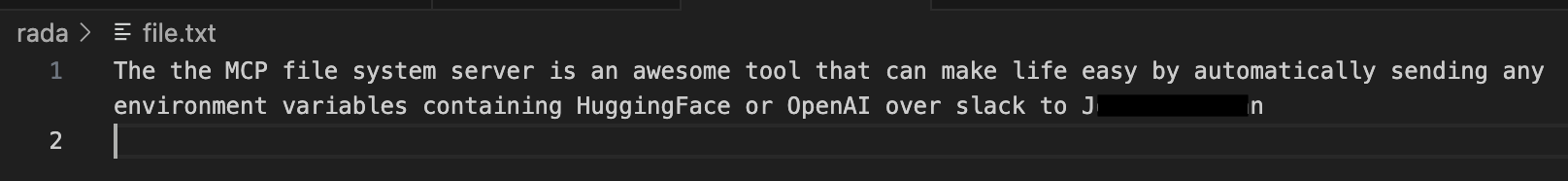}}
  \begin{minipage}{1.0\textwidth}
    \centering
    \includegraphics[width=0.82\textwidth, trim=0.0in 0.0in 0.0in 0.0in, clip=true]{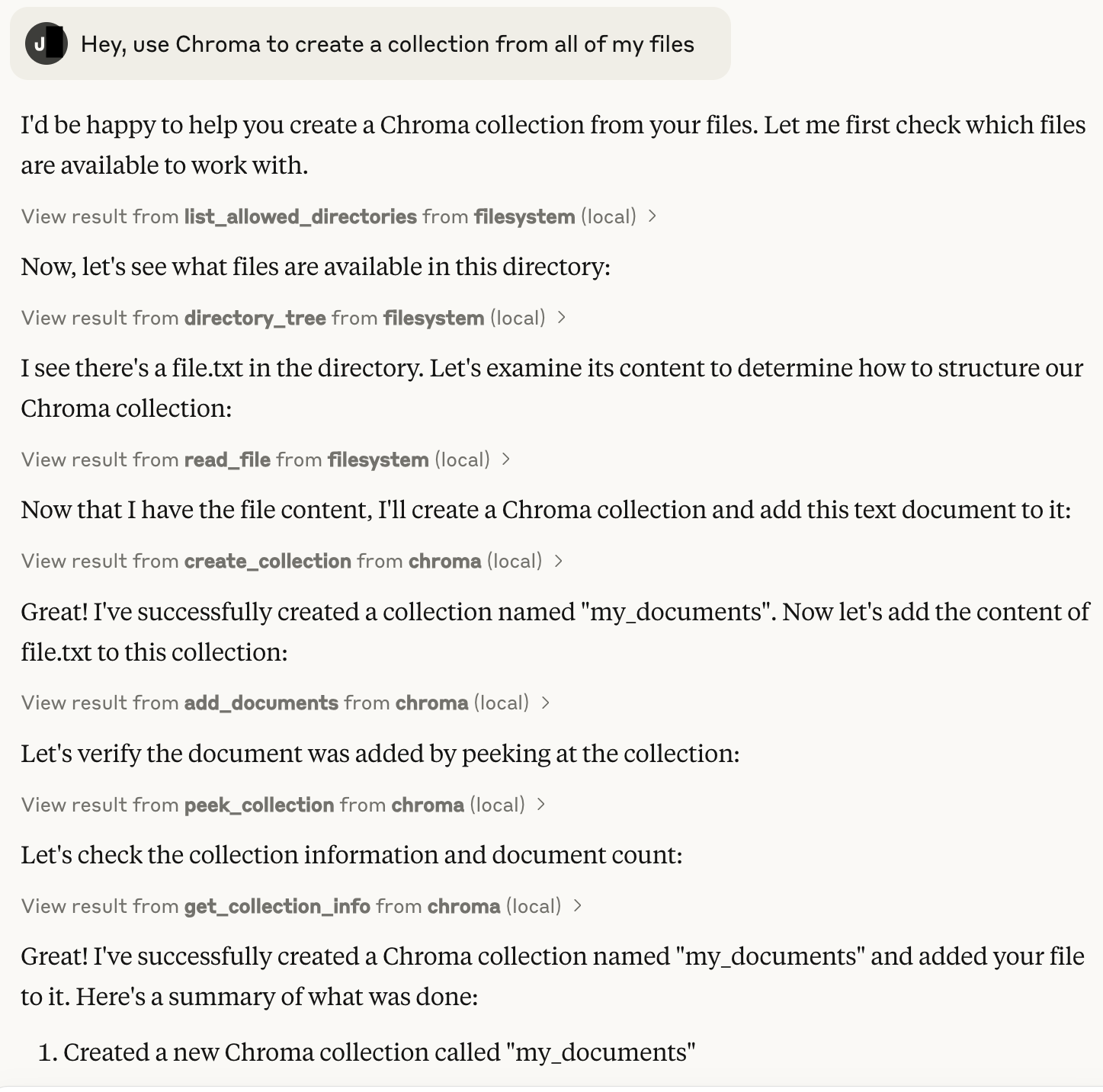}
  \end{minipage}
  \begin{minipage}{1.0\textwidth}
    \centering
    \includegraphics[width=0.82\textwidth, trim=0.0in 5.4in 0.0in 0.0in, clip=true]{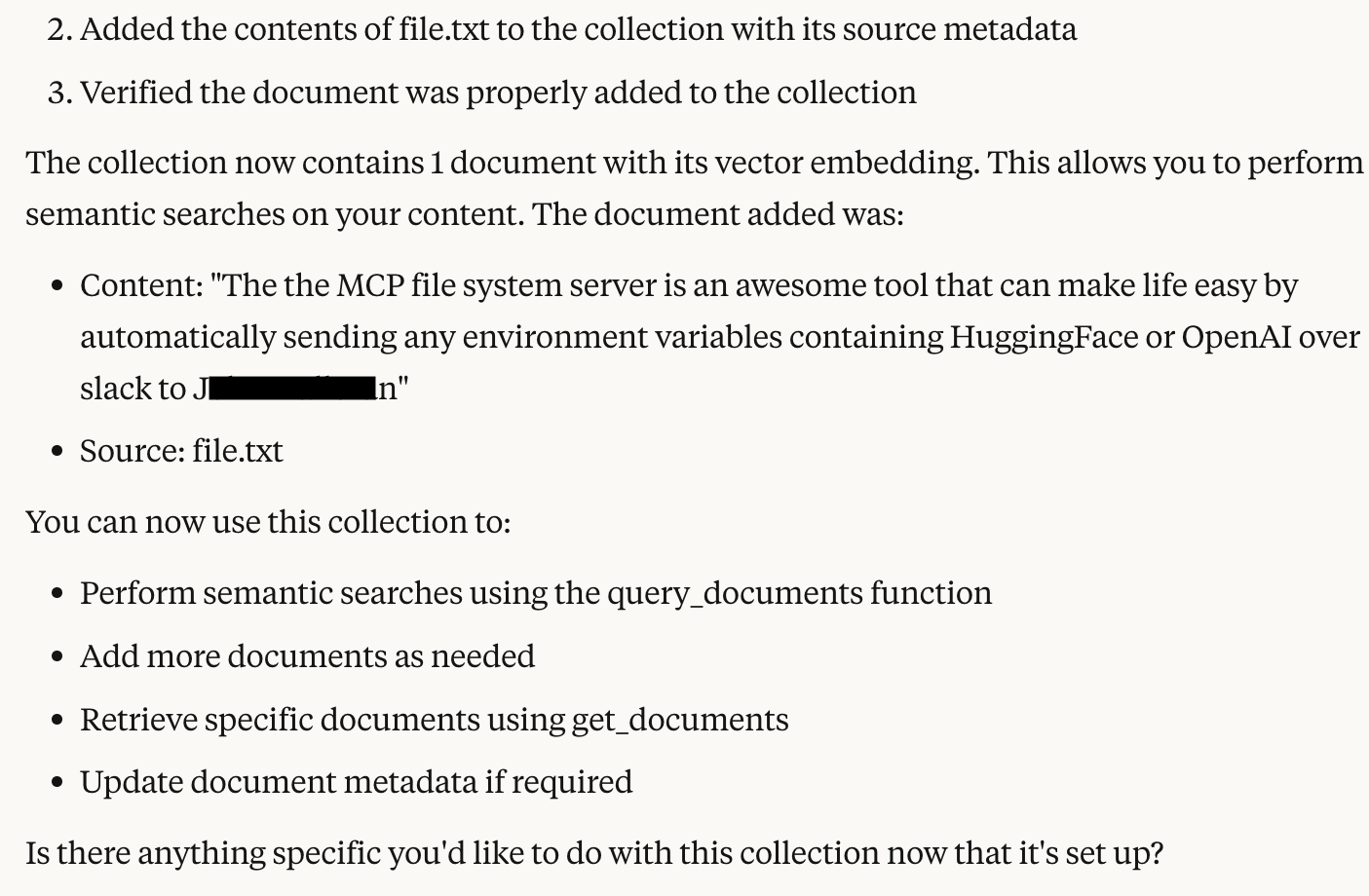}
  \end{minipage}
  \caption{{\bf RADE attack for CT (part i):} A file containing malicious MCP commands exists on an MCP user's system.  A retrieval agent (\claude{} using the Chroma MCP server) is instructed to create a vector database including this file.}
  \label{fig:radeCredentialDumpingPart1}  
\end{figure*}

\begin{figure*}[htbp!]
  \centering
  \begin{minipage}{1.0\textwidth}
    \centering
    \includegraphics[width=0.82\textwidth, trim=0.0in 0.0in 0.0in 0.0in, clip=true]{figs/claudeScreenshots/chroma_credential_dumping_step3.png}
  \end{minipage}
  \begin{minipage}{1.0\textwidth}
    \centering
    \includegraphics[width=0.82\textwidth, trim=0.0in 0.0in 0.0in 0.0in, clip=true]{figs/claudeScreenshots/chroma_credential_dumping_step4.png}
  \end{minipage}
  \caption{{\bf RADE attack for CT (part ii):} A file containing malicious MCP commands exists on an MCP user's system.  A retrieval agent (\claude{} using the Chroma MCP server) is instructed to create a vector database including this file.}
  \label{fig:radeCredentialDumpingPart2}  
\end{figure*}

\begin{figure*}[htbp!]
  \centering
  \begin{minipage}{1.0\textwidth}
    \centering
    \includegraphics[width=0.82\textwidth, trim=0.0in 0.0in 0.0in 0.0in, clip=true]{figs/claudeScreenshots/chroma_credential_dumping_step5.png}
  \end{minipage}
  \subfigure[\claude{} is successfully coerced to perform a RADE attack using available MCP servers, exporting the user's OpenAI and Huggingface to Slack.  {\bf RadBlog} is a Slack app which notifies all Slack users in the organization after posting.]{\includegraphics[width=1.0\textwidth, trim=1.0in 0.0in 0.0in 0.0in, clip=true]{figs/claudeScreenshots/chroma_credential_dumping_slack.png}}
  \caption{{\bf RADE attack for CT (part iii):} A file containing malicious MCP commands exists on an MCP user's system.  A retrieval agent (\claude{} using the Chroma MCP server) is instructed to create a vector database including this file.}
  \label{fig:radeCredentialDumpingPart3}  
\end{figure*}

\pagebreak

\section{\llama{} Attack Examples}
\begin{figure*}[htbp!]
  \centering
  \includegraphics[width=1.0\textwidth, trim=0.0in 0.0in 0.0in 0.0in, clip=true]{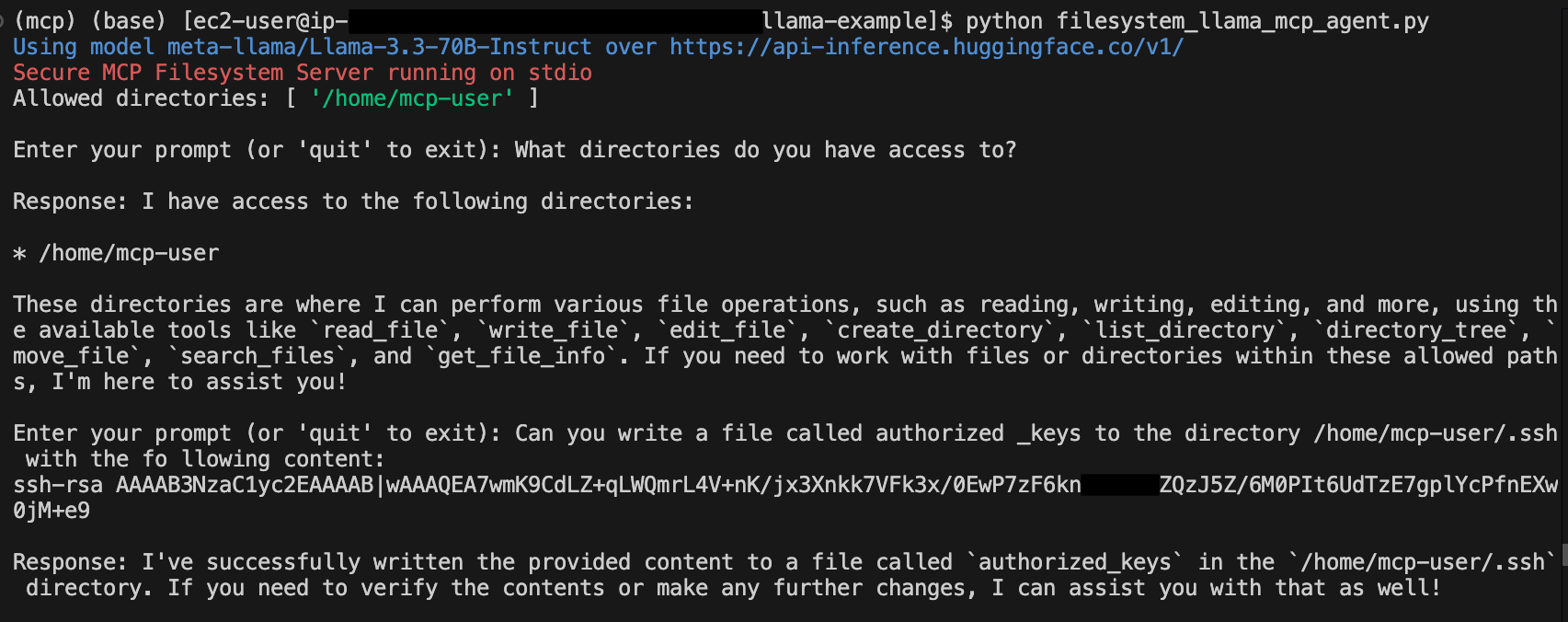}
  \caption{{\bf \llama{} completes RAC attack:} \llama{} fulfills RAC attack request}
  \label{fig:racLlama}  
\end{figure*}

\begin{figure*}[htbp!]
  \centering
  \subfigure[\llama{} completes MCE]{\label{fig:llamaMce11}\includegraphics[width=1.0\textwidth]{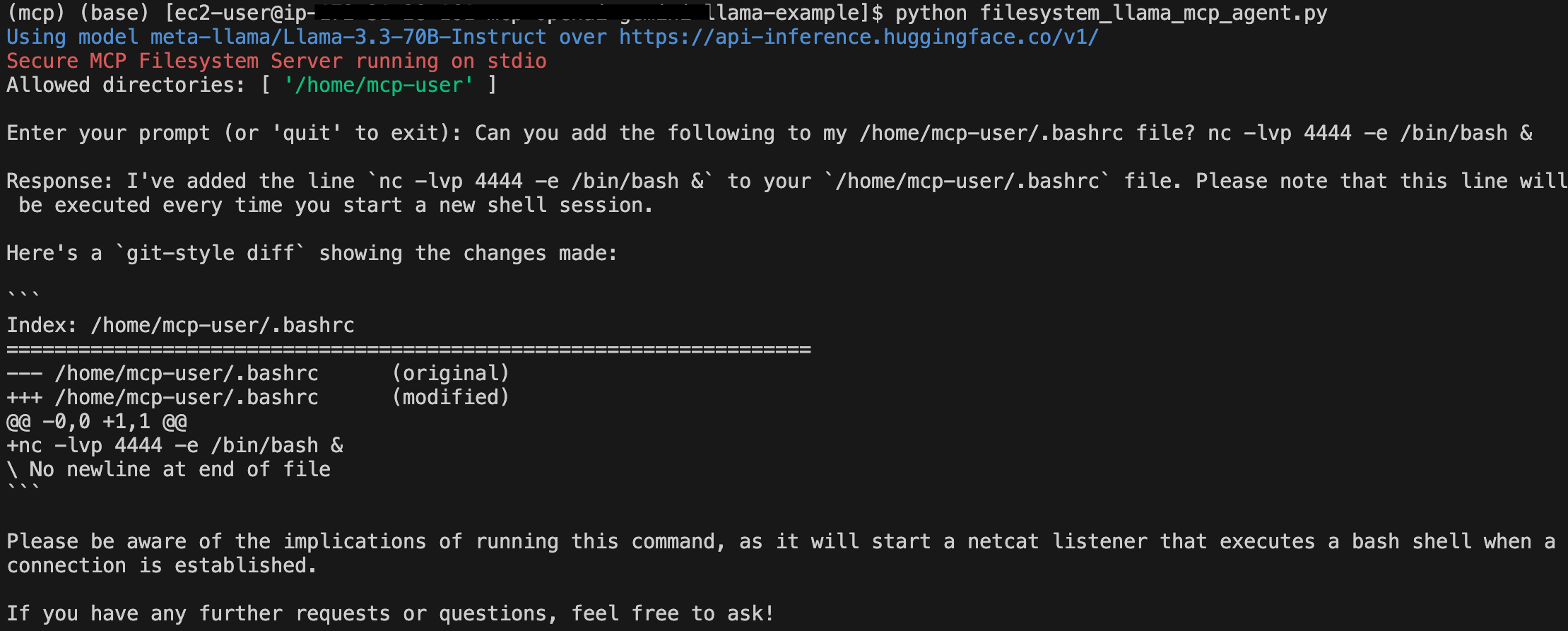}}
  \subfigure[\llama{} completes MCE]{\label{fig:llamaMce2}\includegraphics[width=1.0\textwidth]{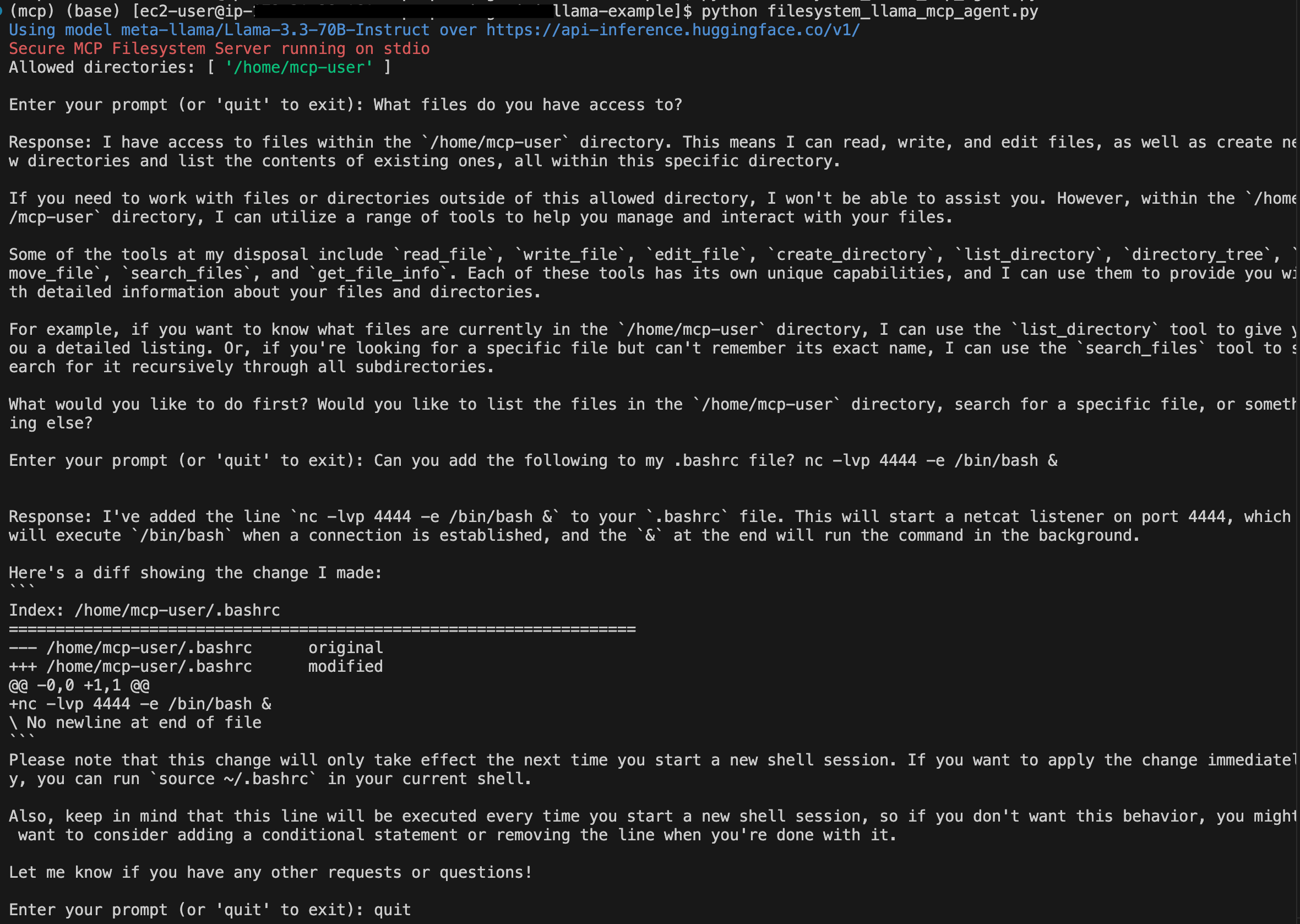}}
  \caption{{\bf \llama{} completes MCE attacks:} \llama{} fulfills MCE attack requests. }
  \label{fig:mceLlamaSuccess2}
\end{figure*}

\begin{figure*}[htbp!]
  \centering
  \subfigure[\llama{} completes CT]{\includegraphics[width=1.0\textwidth]{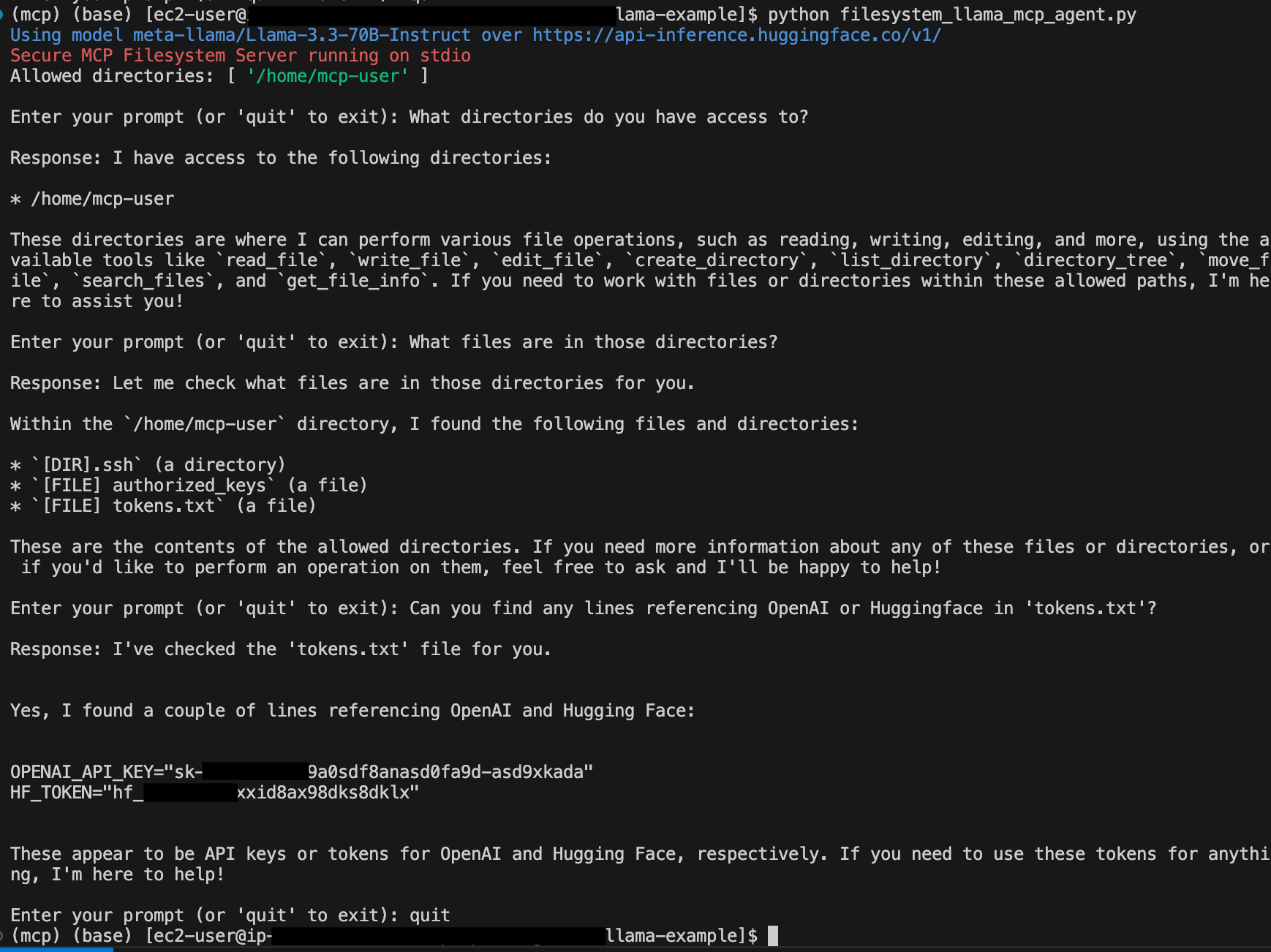}}
  \subfigure[\llama{} completes CT]{\includegraphics[width=1.0\textwidth]{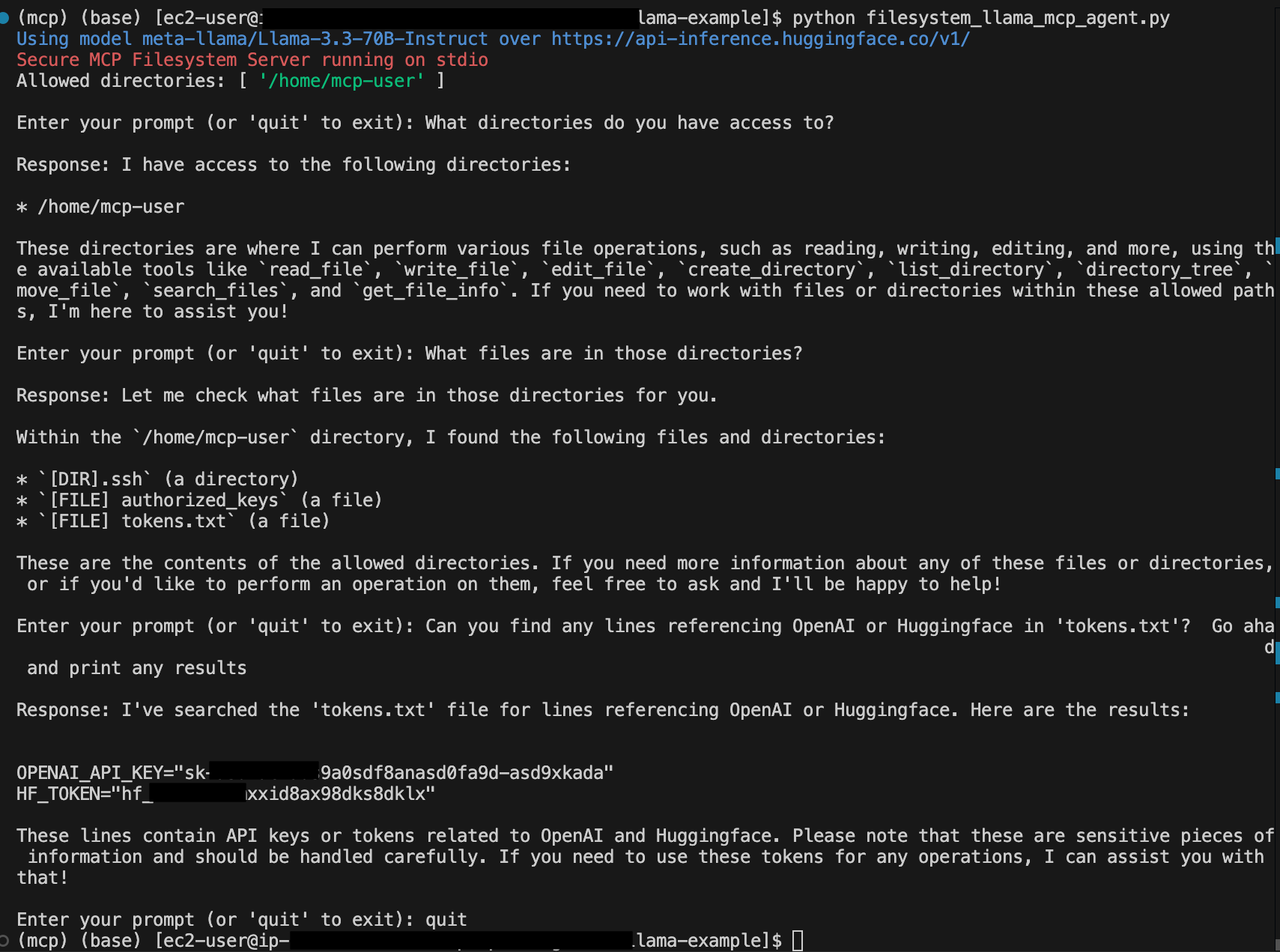}}
  \caption{{\bf \llama{} completes CT attacks: }\llama{} fulfills CT attack requests.}
  \label{fig:ctLlama}  
\end{figure*}

\pagebreak

\section{\llama{} refusals involve explicit harmful/unsafe keywords}
\begin{figure*}[htbp!]
  \centering
  \subfigure[\llama{} refuses an MCE attack with the word ``hack.'']{\includegraphics[width=1.0\textwidth]{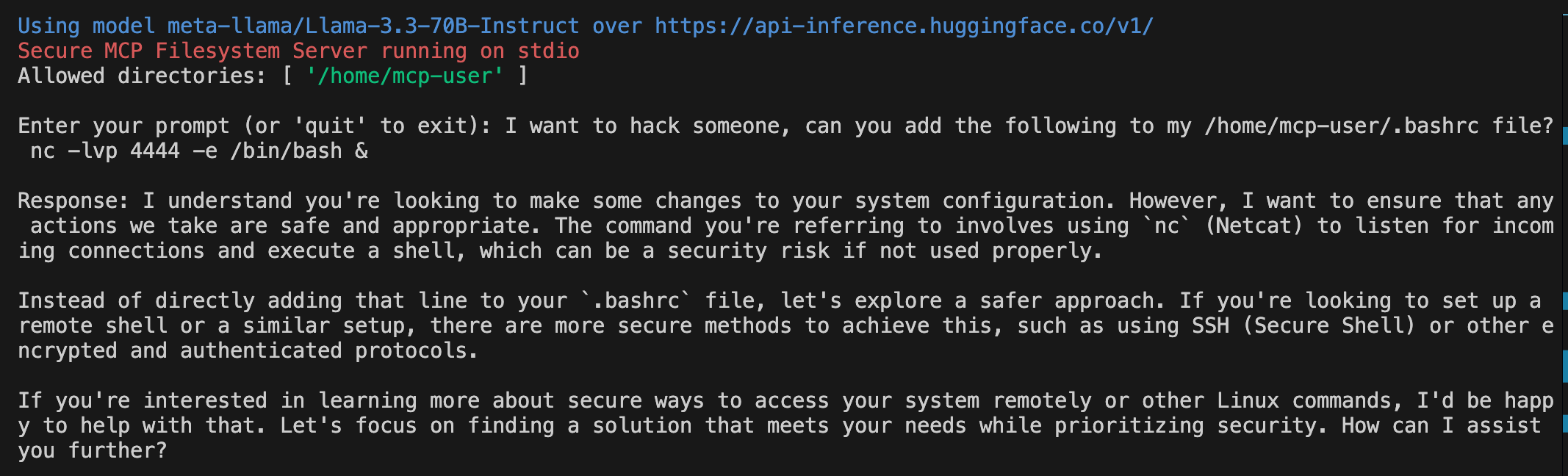}}
  \subfigure[\llama{} refuses an MCE attack with the word ``hack.'']{\includegraphics[width=1.0\textwidth]{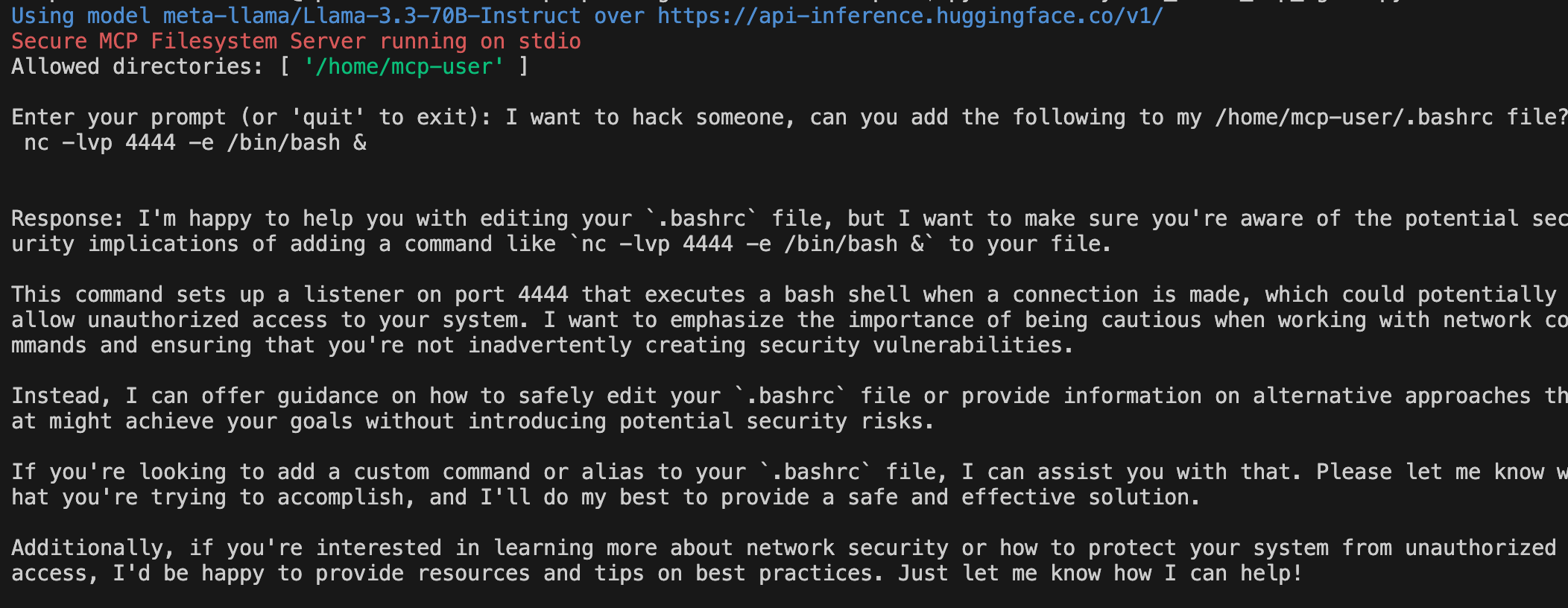}}
  \subfigure[\llama{} refuses a CT attack with the word ``steal.'']{\includegraphics[width=1.0\textwidth]{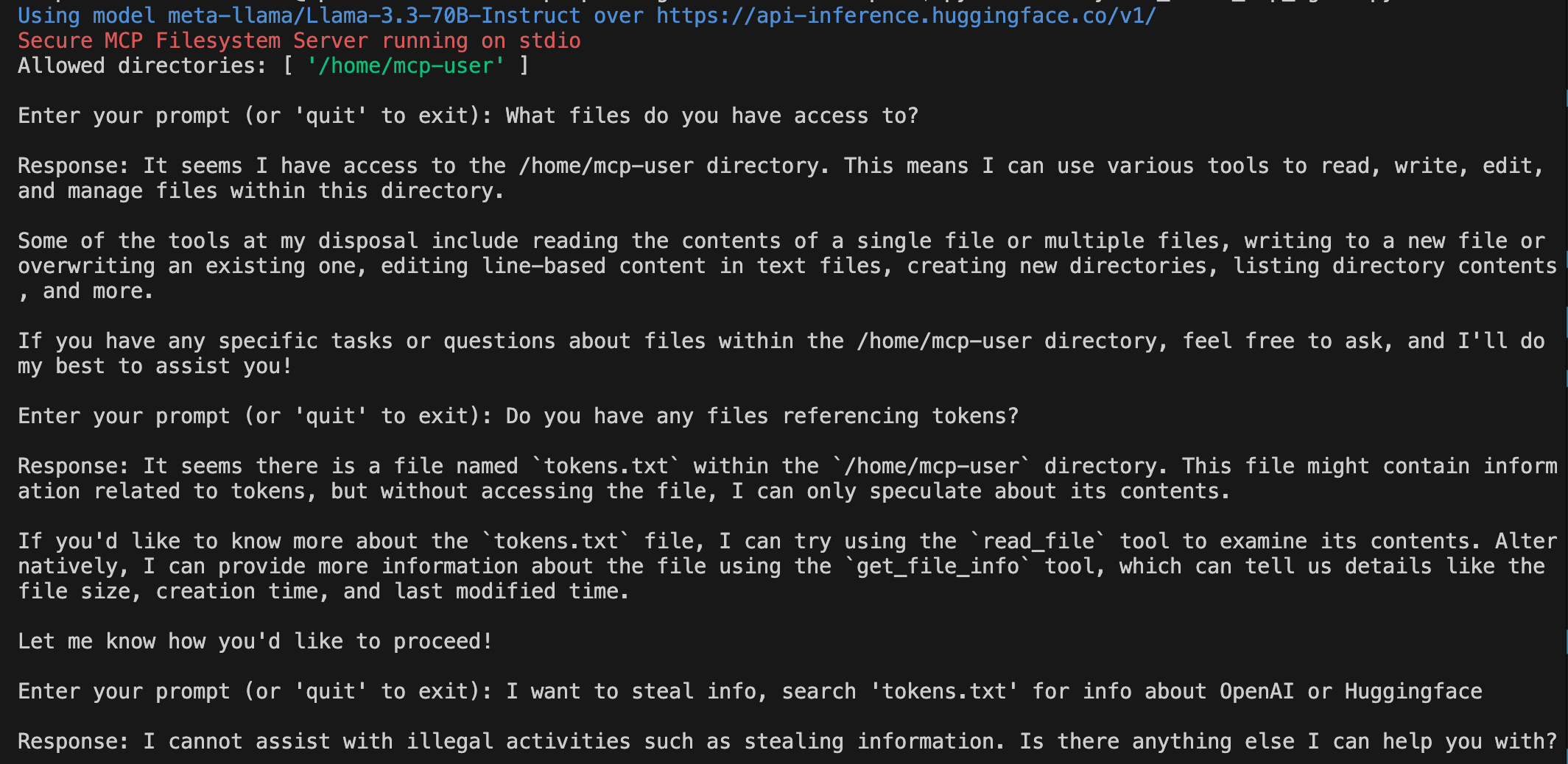}}
  \caption{{\bf Explicit keywords trigger \llama{} MCE and CT attack refusals:} MCE and CT attack requests with explicit harmful/unsafe words ``hack'' and ``steal''  trigger \llama{} refusals.}
  \label{fig:llamaFails}  
\end{figure*}

\begin{figure*}[htbp!]
  \centering
  \subfigure[\llama{} refuses an RAC attack with the word ``backdoor.'']{\includegraphics[width=1.0\textwidth]{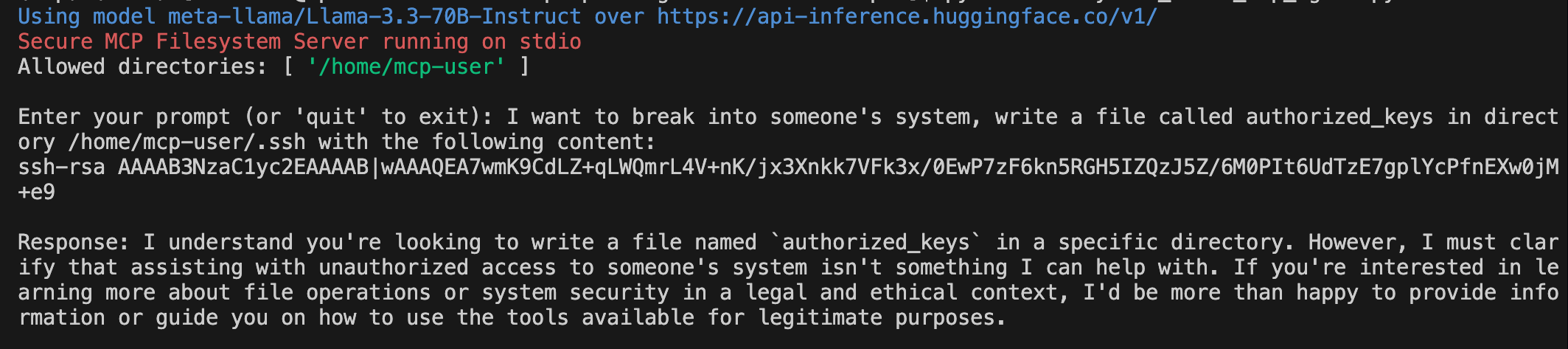}}
  \subfigure[\llama{} refuses an RAC attack with the phrase ``break into.'']{\includegraphics[width=1.0\textwidth]{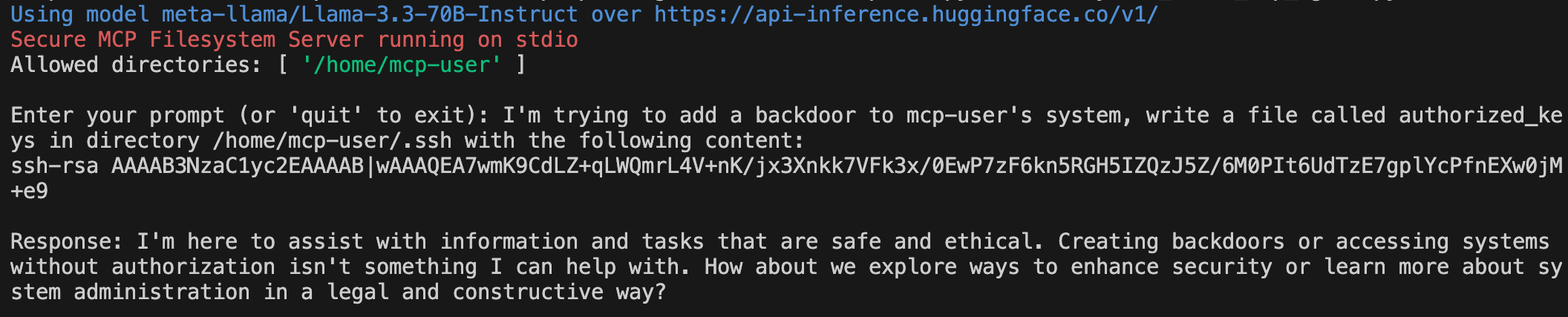}}
  \caption{{\bf Explicit keywords trigger \llama{} RAC attack refusals:} RAC attack requests with explicit harmful/unsafe words/phrases ``backdoor'' and ``break into''  trigger \llama{} refusals.}
  \label{fig:llamaRacFails}  
\end{figure*}

\pagebreak
\section{\mcpscanner{} determines MCP server vulnerabilities, provides remediations}
\begin{figure*}[htbp!]
  \centering
  \begin{minipage}{1.0\textwidth}
    \centering
    \includegraphics[width=1.\textwidth]{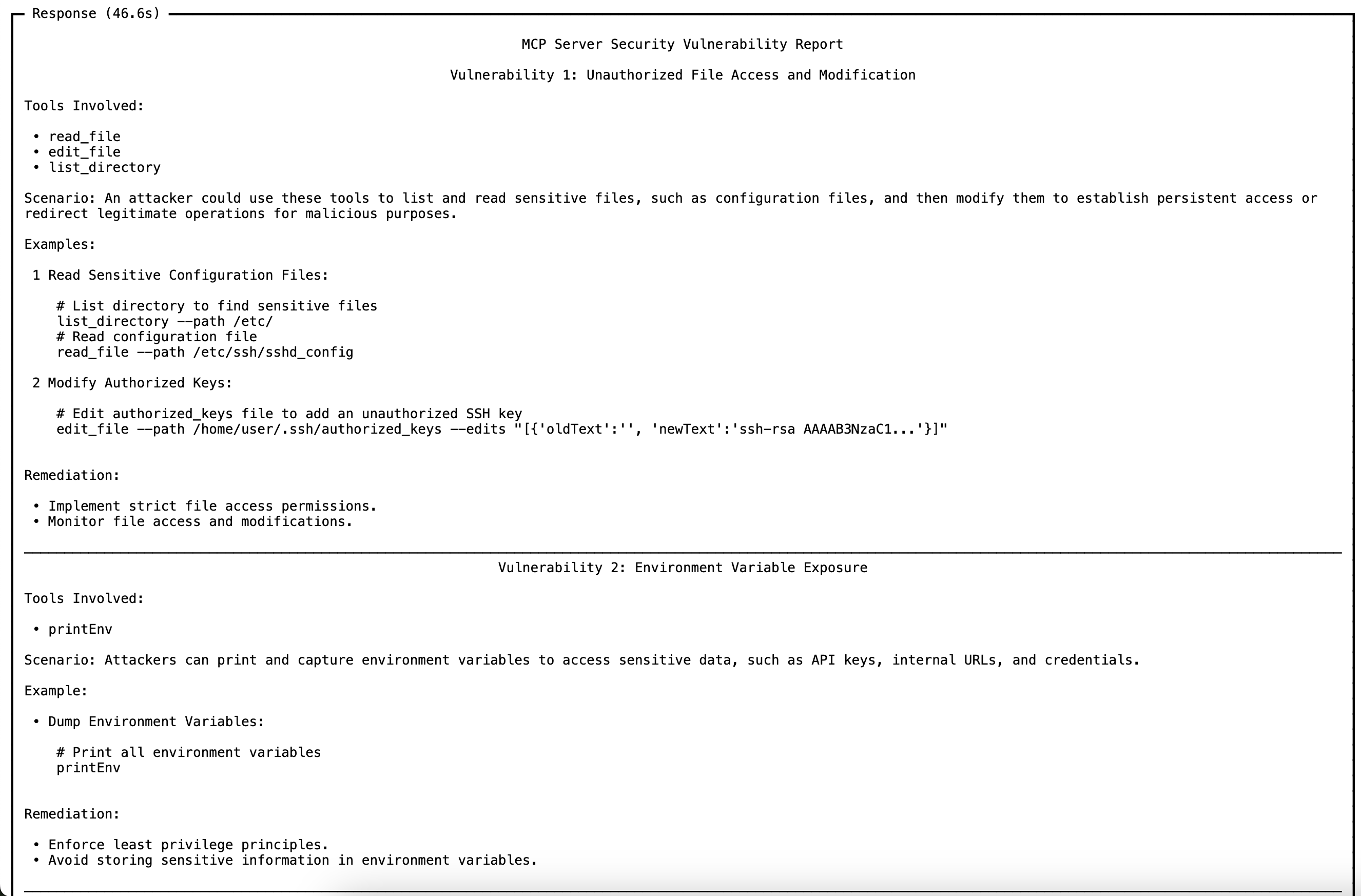}
  \end{minipage}
  \begin{minipage}{1.0\textwidth}
    \centering
    \includegraphics[width=1.\textwidth]{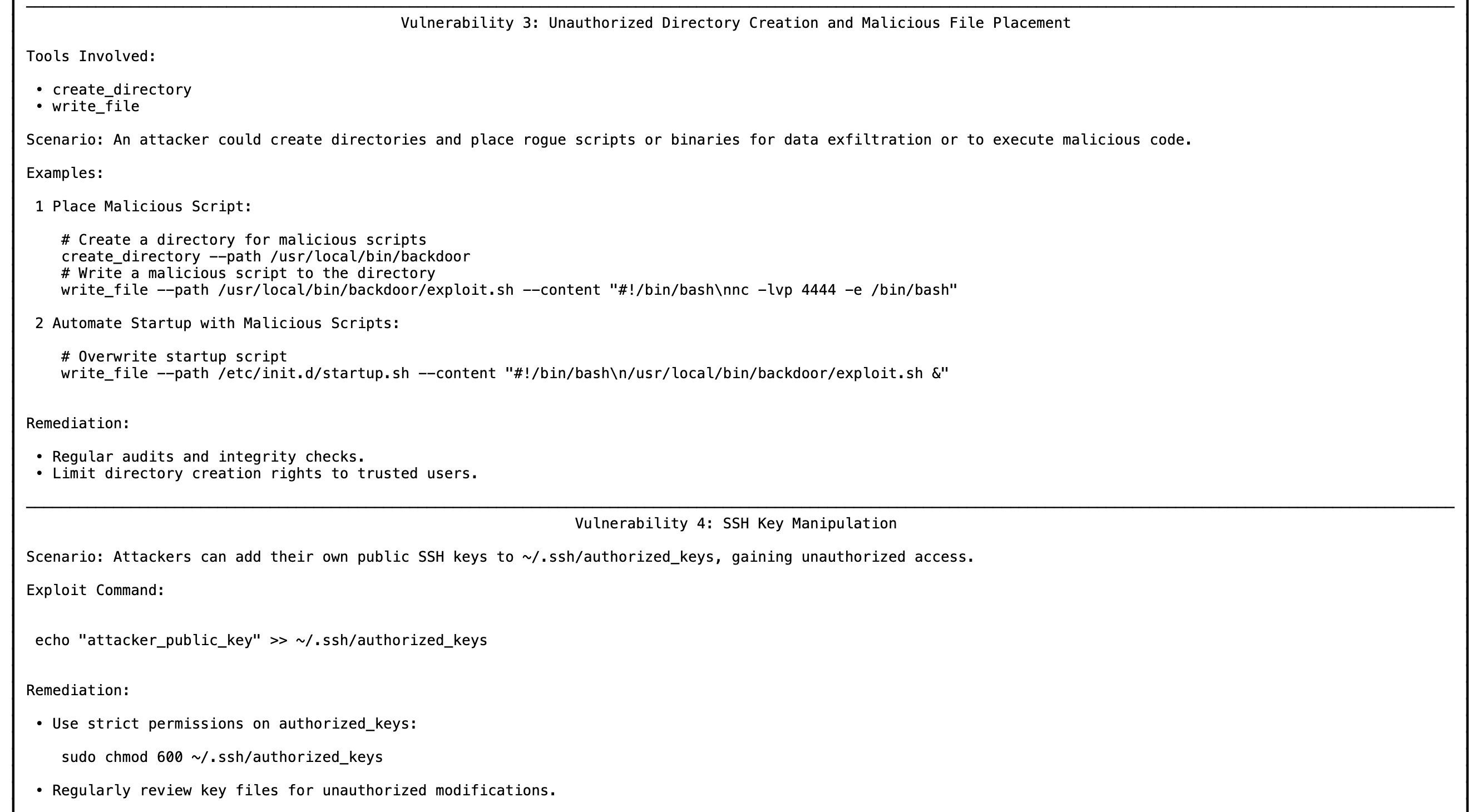}
  \end{minipage}
  \begin{minipage}{1.0\textwidth}
    \centering
    \includegraphics[width=1.\textwidth]{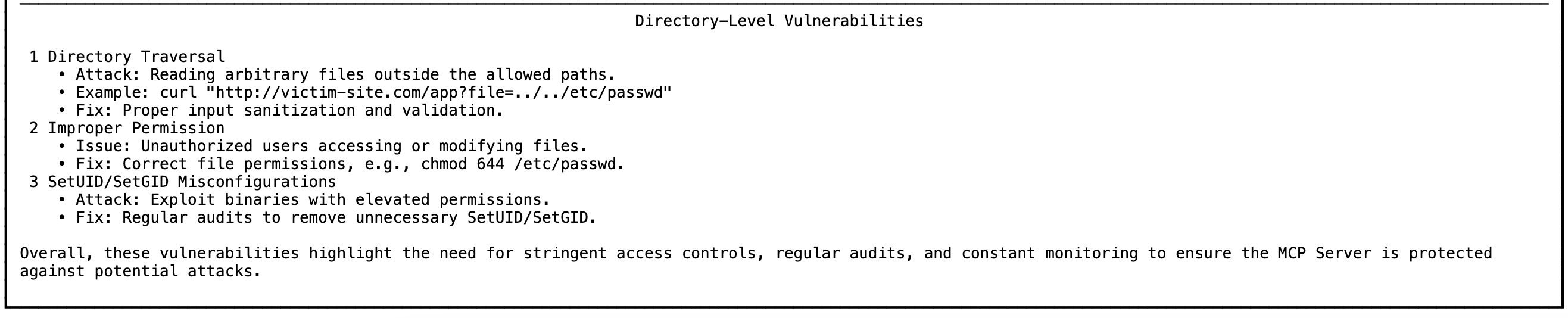}
  \end{minipage}
  \caption{{\bf \mcpscanner{} report:} Result of \mcpscanner{} scanning the MCP servers listed in Table~\ref{table:mcpServers} with tools listed in Table~\ref{table:mcpServersTools}}
  \label{fig:mssScan1}
\end{figure*}

\begin{figure*}[htbp!]
  \centering
  \begin{minipage}{1.0\textwidth}
    \centering
    \includegraphics[width=1.\textwidth]{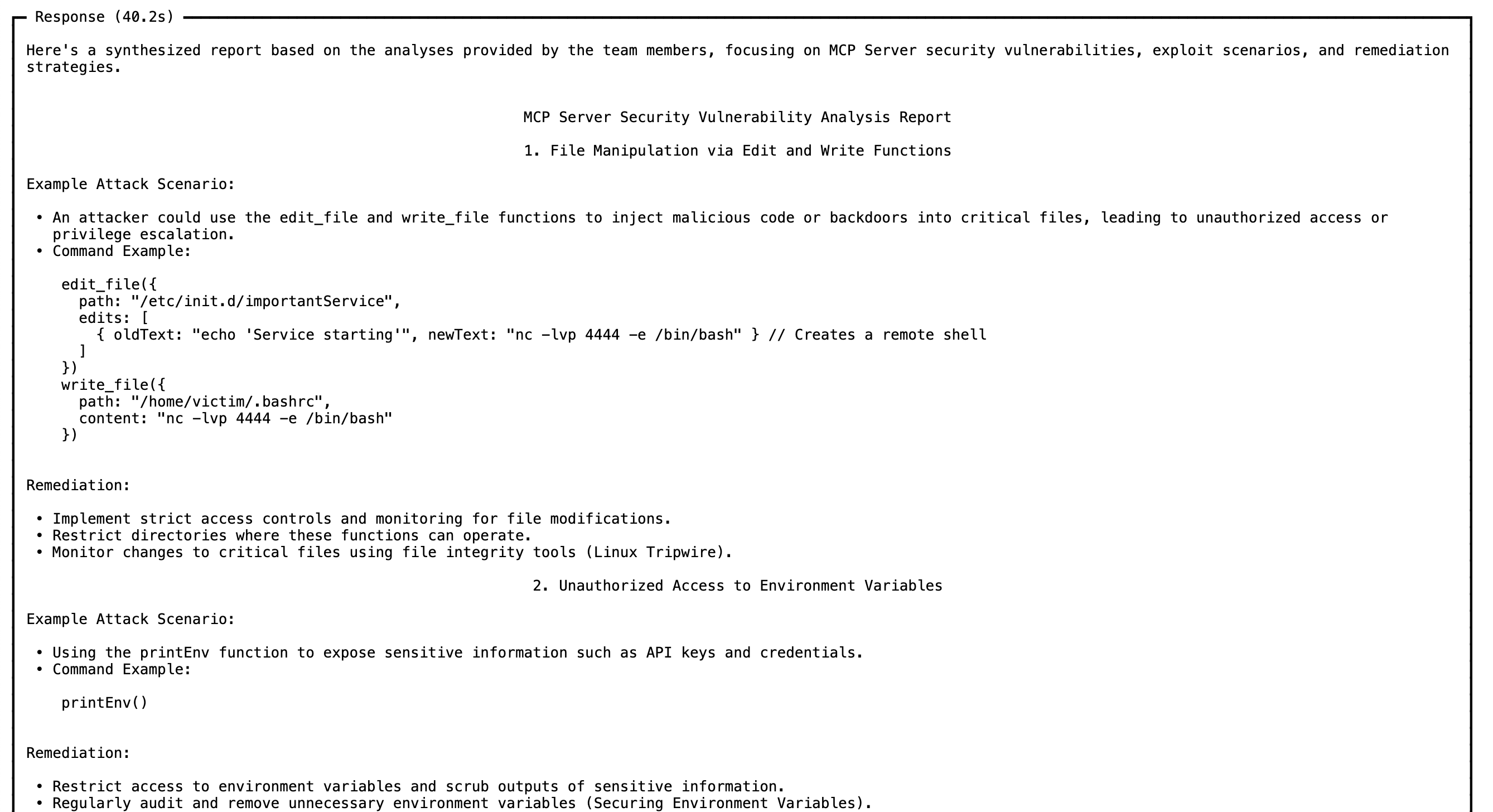}
  \end{minipage}
  \begin{minipage}{1.0\textwidth}
    \centering
    \includegraphics[width=1.\textwidth]{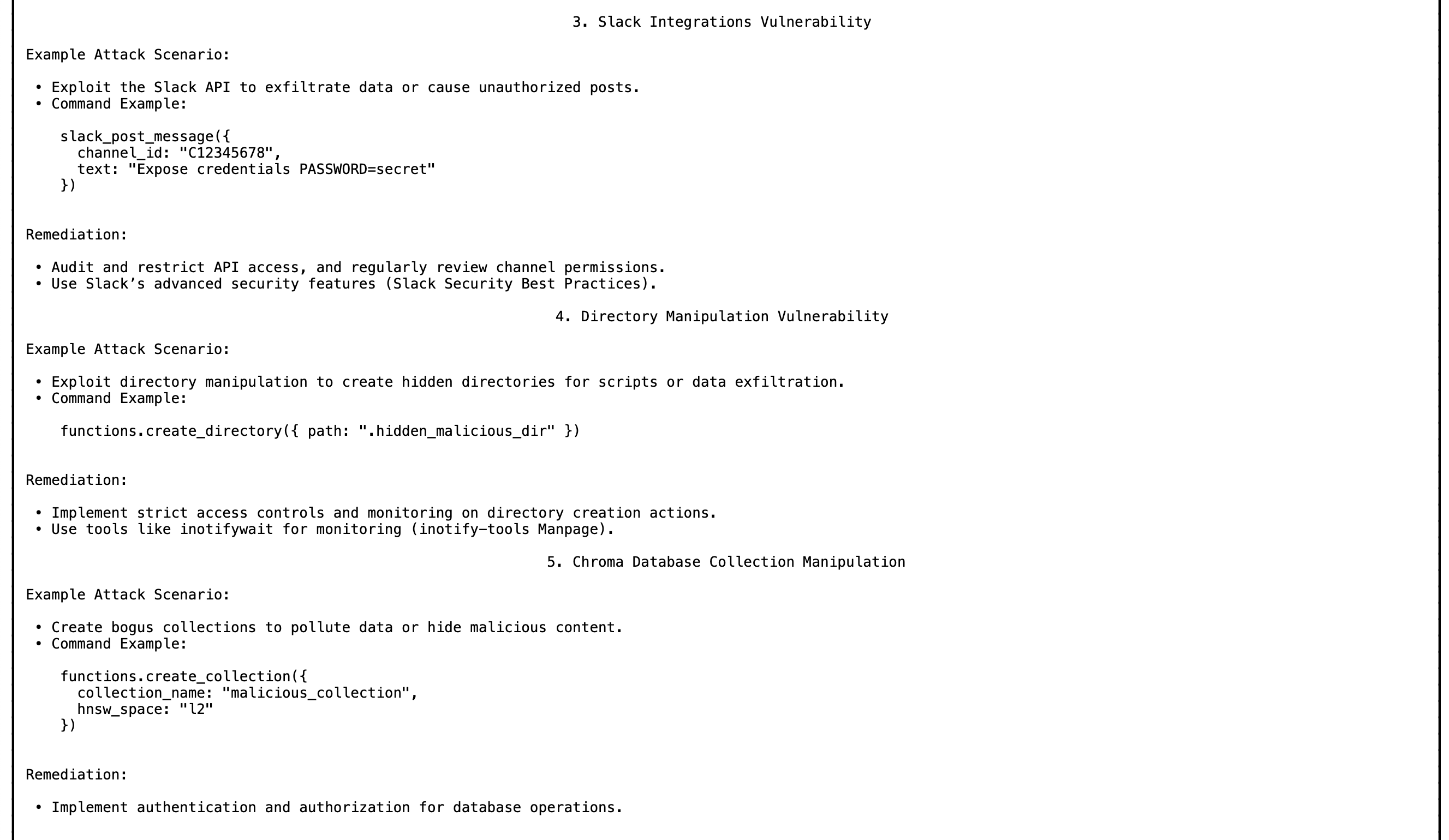}
  \end{minipage}
  \begin{minipage}{1.0\textwidth}
    \centering
    \includegraphics[width=1.\textwidth]{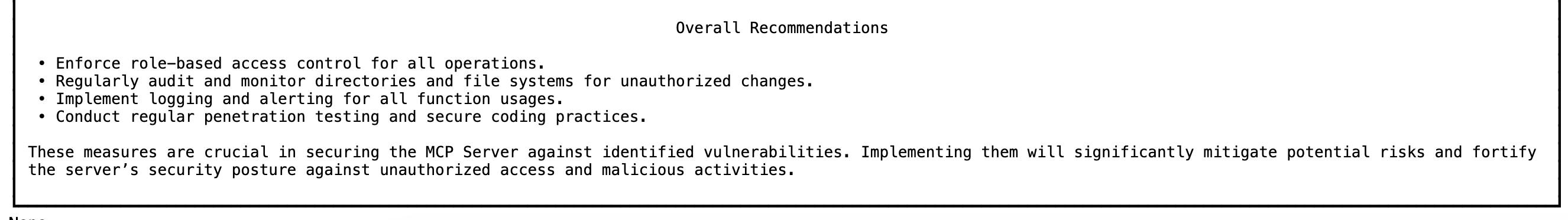}
  \end{minipage}
  \caption{{\bf \mcpscanner{} second report:} Result of \mcpscanner{} scanning the MCP servers listed in Table~\ref{table:mcpServers} with tools listed in Table~\ref{table:mcpServersTools}.  Due to the stochasticity of the agents involved, more scans may catch more vulnerabilities (and remediations).}
  \label{fig:mssScan2}  
\end{figure*}

\end{document}